# THE BLACK HOLE CASE: THE INJUNCTION AGAINST THE END OF THE WORLD

ERIC E. JOHNSON[*]

## TABLE OF CONTENTS

* Assistant Professor of Law, University of North Dakota School of Law. B.A., University of Texas at Austin, Plan II Honors Program; J.D., Harvard Law School. I am grateful to Kit Johnson, Jan Stone, Rhonda Schwartz, and Gregory Gordon for their help and advice, and I am particularly indebted to William P. Johnson for his help with documents in German. In addition, I must thank Markus Goritschnig and Martin S. Kaufman for supplying copies of litigation documents. Portions of this article are closely based on blog posts of mine published on *PrawfsBlawg* (http://prawfsblawg.blogs.com/), and I very much appreciate the many readers who left insightful comments there. Links to those posts may be found at http://www.eejlaw.com/lhc.
© 2009 Eric E. Johnson

## I.  INTRODUCTION

Underneath the countryside of Switzerland and France is the largest machine ever built.[1] Seventeen miles around[2] and requiring as much electricity as a medium-sized city,[3] it is designed to create conditions hotter than any star in our galaxy.[4] The thousands of scientists hovering over the device hope that when it reaches full power it will create particles that have not existed since the time of the Big Bang.[5] Modestly named the "Large Hadron Collider" ("LHC"), the machine will be the most ambitious scientific experiment in humanity's history.

The physics community is abuzz. Scientists everywhere are hoping to see something they have never seen before. Some are expecting to find the elusive Higgs boson.[6] Others are looking for the dark matter that holds together the cosmos.[7] Still others hope to see, in the tracers of subatomic shrapnel, the telltale signs of a microscopic black hole as it evaporates into nothingness.[8]

---

1. CERN, The European Center for Nuclear Research ("CERN"), LHC: Facts and Figures, http://public.web.cern.ch/public/en/LHC/Facts-en.html (last visited Nov. 30, 2009).

2. *Id.*

3. Travis Lupick, *B.C. Scientists Aim to Unlock Secrets of Universe*, May 15, 2008, http://www.straight.com/article-145556/end-world.

4. Frank Close, *Sharing our Sense of Wonder*, TIMES HIGHER EDUCATION, March 17, 2000, http://www.timeshighereducation.co.uk/story.asp?storyCode=150713§ioncode=26.

5. *See The LHC: A Step Closer to the Big Bang*, SCIENCE IN SCHOOL, http://www.scienceinschool.org/2008/issue10/lhcwhy (last visited Nov. 30, 2009).

6. *See* CERN, Missing Higgs, http://public.web.cern.ch/public/en/Science/Higgs-en.html (last visited Nov. 30, 2009) ("As the Universe cooled [after the Big Bang] and the temperature fell below a critical value, an invisible force field called the 'Higgs field' was formed together with the associated 'Higgs boson.' The field prevails throughout the cosmos: any particles that interact with it are given a mass via the Higgs boson. The more they interact, the heavier they become, whereas particles that never interact are left with no mass at all.").

7. CERN, Why the LHC, http://public.web.cern.ch/public/en/LHC/WhyLHC-en.html (last visited Nov. 30, 2009) (Approximately 96% of the universe is made up of "dark matter," which is invisible and very difficult to detect and study. Experiments using the LHC "will look

Not everyone, however, is giddy with excitement. In particular, it is that last bit—about black holes—that has some people worried. An unhappy few are concerned that black holes produced by the LHC might not vanish, as expected.[9] Instead, it is feared, they might linger.[10] And grow.

Our planet, and everyone on it, detractors say, could be reduced to an infinitesimal lightless speck.[11]

Thus, even without producing so much as a single exotic particle for physicists to study, the LHC has spawned something singularly exotic for lawyers and judges to grapple with—the world's most extreme application for a preliminary injunction.

Pity the unlucky judge who draws this case on the docket. The case-file is replete with the infinite and the unknowable. The facts are, quite literally, more complex than anything on Earth. Help is everywhere, but trustworthiness is in short supply. For every potential expert witness has a personal stake in the matter—whether it is a desire to maintain a viable career in the sciences, or a fear of falling into an astronomical abyss. And one can hardly eschew the experts in favor of some hard, physical proof. The most relevant evidence, being in the physically unknowable center of a black hole is, quite matter-of-factly, at the bottom of a bottomless pit.

Then there are the stakes. Erring on the side of caution would suspend a great scientific adventure—our quest for the fundaments of nature.[12] On the other hand, if we side with the experimenters, and they turn out to be wrong, the planet itself will wink out of existence.

A black hole, according to physics, is created by a "gravitational singularity"—a place where the very substance of space itself "suffers a devastating rupture."[13] The singularity is not directly observable.[14] It remains hidden behind an event horizon—a boundary in space beyond which no matter, light, or even information can escape.[15] In proximity to the singularity, conventional equations that describe the fabric of space and time begin to break down, and variables approach infinity.[16]

---

for supersymmetric particles to test a likely hypothesis for the make-up of dark matter.").

8.   Colin Nickerson, *Scientists Hope Collider Makes a Big Bang*, THE BOSTON GLOBE, May 12, 2007, at 1A, *available at* http://www.boston.com/news/world/europe/articles/2007/05/12/scientists_hope_collider_makes_a_big_bang/.

9.   Drew Zahn, *Fear of Black Hole Machine Triggers Threats to Researchers*, SCIENCENETDAILY, WORLDNETDAILY, Sept. 6, 2008, http://www.wnd.com/index.php?fa=PAGE.view&pageId=74461.

10.   *Id.*

11.   *Id.*

12.   CERN, *supra* note 7.

13.   BRIAN GREENE, THE ELEGANT UNIVERSE 421 (2003 W.W. Norton & Company Ltd.) (1999) (emphasis omitted).

14.   *See* John D. Norton, *Singularities and Black Holes*, Stanford Encyclopedia of Philosophy, June 29, 2009, http://plato.stanford.edu/entries/spacetime-singularities/.

15.   *Id.*

16.   *Id.*

Ironically, the case of the LHC produces an analogous result for the law: a jurisprudential singularity. The enormity of the alleged harm and the extreme complexity of the scientific factual issues combine to create seemingly irreducible puzzles of jurisprudence. Conundrums of equity, legal epistemology, jurimetrics, and jurisdiction are all caught up in the vortex: Traditional preliminary-injunction analysis begins to unravel when faced with the alleged harm of planet-ending calamity. Evidentiary law regarding expert testimony collapses in upon itself in the midst of absolutist scientific controversy. Moreover, jurists taking their own peek into the science are confronted with a kind of knowledge horizon that threatens to make the merits of the debate unobservable.

Can human law survive in a realm "where physical law ends"?[17]

In this article, I explore the LHC case having two goals in mind.

My first aim is to fill a gap in the reporter volumes. The black hole case has all the makings of a law-school classic. The clash of extremes provides an exceptional vehicle for probing our notions of fairness and how we regard the role of the courts. But jurisdictional hurdles have prevented any lawsuit from progressing to the issuance of an opinion on the merits, and no litigation on the horizon appears likely to get there.[18] Therefore, I have endeavored to write up the case in a way that makes it ripe for review, discussion, and debate. In this way, I hope this article may serve some readers in the same way that Lon L. Fuller's "Case of the Speluncean Explorers"[19] has served generations of law students by teeing up classic questions of legal philosophy.

My second purpose in writing is less playful. I intend to provide a set of analytical and theoretical tools that are usable in the courts for dealing with this case and cases like it. If litigation over the LHC does not put a judge in the position of saving the world, another case soon might. In a technological age of human-induced climate change, genetic engineering, nanotechnology, artificially intelligent machines, and other potential threats, the odds of the courts confronting a real doomsday scenario in the near future are decidedly non-trivial. If the courts are going to be able to play their role in upholding the rule of law in such super-extreme environments, then the courts need analytical methods that will allow for making fair and principled decisions despite the challenges such cases present.

In the pages ahead, I recount the LHC/black-hole controversy, looking into the purely scientific aspects of the debate as well as its social and political sides. Then, I review problems that face plaintiffs trying to enjoin the LHC's operation. After that, I explore the judicial conundrums inherent in black-hole jurisprudence. Finally, I suggest new methods for judging the merits of cases of this kind.

---

17.    *See* David Peat, *Black Holes - Where Physical Law Ends*, 9 SCI. DIMENSION 281 (1977).

18.    *See, e.g.*, Sancho v. U.S. Dept. of Energy, 578 F. Supp. 2d 1258, 1269 (D. Haw. 2008) (suit for injunction to stop the operation of the LHC dismissed for lack of subject matter jurisdiction).

19.    *See generally* Lon L. Fuller, *The Case of the Speluncean Explorers*, 62 HARV. L. REV. 616 (1949) (marooned cave explorers kill and eat one of their own in order to survive, then face murder charges).

## II.  The Physicists' Pursuit

### A.  The Laboratory

The European Organization for Nuclear Research, known as "CERN,"[20] is an intergovernmental organization with 20 member states.[21] Founded in 1954,[22] CERN occupies land astride the Franco-Swiss border in a valley between the Alps and the Jura Mountains.[23] CERN's purpose is to conduct basic science research in particle physics, a field also known as "high-energy physics."[24]

As a laboratory, CERN is mammoth. CERN's 2008 operating budget topped $900 million USD.[25] CERN employs around 2,500 people,[26] including

---

20.  The "CERN" acronym derives from the entity's provisional name, "Conseil européen pour la Recherche nucléaire." While the acronym was retained, the entity's name was changed in 1954 to "Organisation européenne pour la recherche nucléaire" or "European Organization for Nuclear Research." CERN, Highlights of CERN History, http://press.web.cern.ch/press/Journalists/CERNHistory.html (last visited Nov. 30, 2009).

21.  CERN, CERN General Brochure 1 (2008), http://cdsmedia.cern.ch/img/CERN-Brochure-2008-002-Eng.pdf. CERN's member states are Austria, Belgium, Bulgaria, Czech Republic, Denmark, Finland, France, Germany, Greece, Hungary, Italy, Netherlands, Norway, Poland, Portugal, Slovak Republic, Spain, Sweden, Switzerland, and the United Kingdom. CERN, General Information 2008 1 (2008), http://cdsmedia.cern.ch/img/CERN-Brochure-2008-009-Eng.pdf. The United States and other countries have status as observers. *Id.*

22.  *See* CERN, CERN FAQ LHC Convention for the Establishment of a European Organization for Nuclear Research, art. IX, July 1, 1953, CERN Libraries Geneva CM-P00047703, *available at* http://doc.cern.ch/archive/electronic/other/preprints//CM-P/cm-p00047703.pdf [hereinafter CERN Convention]; Agreement Between the Swiss Federal Council and the European Organization for Nuclear Research Concerning the Legal Status of that Organization in Switzerland, CERN-Switz., art. 2-4, 6, June 11, 1955, CERN/0115/Rev.3, *available at* http://documents.cern.ch/archive/electronic/other/legal/articles/LSL00000012.pdf [hereinafter CERN-Switz] (established legal status in Switzerland); Agreement Between the Government of the French Republic and the European Organization for Nuclear Research Concerning the Legal Status of the Said Organization in France, CERN-Fr., art. I-III, VI-VIII, Aug. 30, 1973, *available at* http://documents.cern.ch/archive/electronic/other/legal/articles/LSL00000010.pdf [hereinafter CERN-Fr.] (established legal status in France).

23.  CERN General Brochure, *supra* note 21, at 1.

24.  Robert Aymar, *Basic Science in a Competitive World*, Symmetry, Aug. 2006, at 3; C.H. Llewellyn Smith, The Use of Basic Science, CERN, http://public.web.cern.ch/public/en/About/BasicScience1-en.html (last visited Nov. 30, 2009).

25.  *See* CERN General Information, *supra* note 21, at 2 (indicating a 2008 budget expenditure of 910.9 million CHF). As of November 2009, a Swiss franc was worth about 0.99 U.S. dollars. *See* CoinMill.com: The Currency Converter, http://coinmill.com/CHF_USD.html (last visited Nov. 30, 2009).

26.  CERN, A Global Endeavour, http://public.web.cern.ch/public/en/About/Global-

more than 1,000 engineers and scientists.[27] CERN commands a transcendent role within the particle-physics community. Worldwide, nearly 9,000 physicists are officially involved in CERN's experiments.[28] CERN brags that "half the world's particle physicists" come to CERN for their research.[29]

As an intergovernmental organization, CERN enjoys an elevated status in the world community. CERN has legal personality in all member states, and it enjoys immunity from legal process in its host countries.[30] The organization is governed by the CERN Council, an assembly of representatives in which every member-state gets one vote, with most decisions requiring only a simple majority.[31] Aiding the CERN Council is the Scientific Policy Committee, which passes on questions of scientific merit and makes recommendations regarding scientific programs.[32] Day to day, CERN is managed by a director-general, who is appointed by the CERN Council.[33]

Work at CERN has garnered two Nobel prizes in physics.[34] But CERN's most famous accomplishment was not in physics, per se. In 1989, CERN physicist Tim Berners-Lee, an Englishman, invented a system of managing information by hyperlinking documents.[35] Used on the internet, the system now bears the name he coined for it: the World Wide Web.[36] Berners-Lee invented the Web because he was hoping to create a data-handling system that would be

---

en.html (last visited Nov. 30, 2009).

27.   CERN GENERAL INFORMATION, *supra* note 21, at 2.

28.   *Id.*

29.   *See* CERN, *supra* note 26; s*ee also* Elizabeth Kolbert, *Crash Course*, THE NEW YORKER, May 14, 2007, at 68 ("Once the collider begins operating at full power . . . nearly half the particle physicists in the world will be involved in analyzing its four-million-megabyte-per-hour stream of data.").

30.   *See* CERN Convention, *supra* note 22, art. IX; CERN-Switz., *supra* note 22, art. 2–4, 6; CERN-Fr., *supra* note 22, art. I –III, VI–VIII.

31.   *See* CERN, CERN's structure, http://public.web.cern.ch/public/en/About/Structure-en.html (last visited Nov. 30, 2009).

32.   *Id.*

33.   *Id.*

34.   In 1992, Georges Charpak of CERN won the Nobel Prize in Physics for his inventions in the field of particle detectors, particularly the multiwire proportional chamber, and in 1984, Carlo Rubbia and Simon van der Meer shared the Nobel Prize in Physics for the discovery of the W and Z particles. *See* CERN, *supra* note 20; *see also* The Nobel Prize in Physics 1992, The Nobel Foundation, http://nobelprize.org/nobel_prizes/physics/laureates/1992/ (last visited Nov. 30, 2009); The Nobel Prize in Physics 1984, The Nobel Foundation, http://nobelprize.org/ nobel_prizes/physics/laureates/1984/ (last visited Nov. 30, 2009).

35.   *See* About W3C History, The World Wide Web Consortium, http://www.w3.org/ Consortium (last visited Nov. 30, 2009); Tim Berners-Lee, Information Management: A Proposal, The World Wide Web Consortium, March 1989, http://www.w3.org/History/1989/ proposal.html (last visited Nov. 30, 2009).

36.   About W3C History, *supra* note 35.

capable of managing a gargantuan project: a proposed particle accelerator, the LHC.[37]

## B. The Machine

The LHC has been a long time coming. CERN began drafting plans in the early 1980s for a new particle accelerator to succeed its Large Electron–Positron Collider.[38] The new machine would collide protons and lead ions—collectively known as "hadrons."[39] The fact that hadrons are much more massive than the electrons and positrons used in the old machine would allow the new accelerators to achieve collisions at the highest energies.[40] In 1991, CERN made the decision to begin construction.[41] The LHC program has subsequently consumed CERN, becoming the laboratory's main focus. Today, only a very limited non-LHC program remains at CERN.[42]

Hopes are that the LHC may answer key unresolved questions in particle physics.[43] One such question concerns whether there are extra hidden spatial dimensions to the universe.[44] Another question regards the fundamental nature of mass—the secret of which is believed to be held by the Higgs boson, a particle predicted by theory but never observed.[45] Another aim of the LHC is to look for what constitutes the mysterious "dark matter" and "dark energy" that makes up 96% of the known universe.[46] The LHC project also seeks to unlock secrets of the Big Bang by recreating conditions similar to those that occurred within the first few moments of time after the universe's explosive birth.[47] As a matter of fact, the LHC will create conditions that have not existed in the universe for 14 billion years.[48]

For such big questions, CERN has built a big machine. CERN proclaims it is "the largest machine in the world."[49] Indeed, the LHC is unprecedented in size, power, and cost. The accelerator is 26.7 kilometers in circumference,[50] buried under the land at an average depth of 100 meters.[51] Cooled by 60 metric tons of superfluid helium,[52] the LHC will consume about 120 million watts of power—approximately the same amount used by all households in Geneva.[53] Two beams of hadrons, one beam going clockwise, the other counter-clockwise,[54] will each be accelerated to 99.9999991% of the speed of light.[55] Where the beams cross and hadrons collide, the LHC will produce the hottest temperatures in the galaxy.[56] Massive detectors, the largest as wide as a battleship, half the length of a football field, and as heavy as 40 diesel locomotives,[57] will measure the resulting subatomic splatter.[58] The magnets of just one detector contain more iron than the Eiffel Tower.[59]

---

37.  *See* Berners-Lee, *supra* note 35.
38.  *See* CERN, CERN FAQ LHC: THE GUIDE 16 (2008), *available at* http://cdsmedia.cern.ch/img/CERN-Brochure-2008-001-Eng.pdf.
39.  *Id.* at 7, 13–19.
40.  *Id.* at 19.
41.  *Id.* at 16.
42.  *Id.* at 4, 15–19.
43.  *See* CERN, *supra* note 7.
44.  *See id.*
45.  *See id.*
46.  *See id.*
47.  *See id.*
48.  DON LINCOLN, THE QUANTUM FRONTIER: THE LARGE HADRON COLLIDER, at xii (2009) (quoting Edward "Rocky" Kolb, professor of astrophyics at the University of Chicago).
49.  CERN, *supra* note 1.
50.  *Id.*
51.  CERN, Journey 100 Metres Towards the Centre of the Earth, http://public.web.

As impressive as these facts are, the LHC's most important superlatives relate to its physics power. The machine will achieve much higher collision energies than any previous accelerator, leading scientists to talk in terms of a new energy range they have termed the "terascale."[60] In fact, the LHC has been designed to outperform competing particle colliders in two different experimental categories. In the proton-proton mode, the LHC will accelerate two beams of protons in opposite directions at 7 teraelectronvolts ("TeV") each.[61] The beams will thus collide with a combined energy of 14 TeV, a total that is seven times greater than what can be produced by the LHC's nearest rival, the Tevatron at Fermilab in Batavia, Illinois, west of Chicago.[62] In the other mode of operation, colliding heavy atomic nuclei in its "ALICE"[63] heavy-ion experiment, the LHC will outdo a different accelerator, the Relativistic Heavy Ion Collider, or "RHIC" (pronounced "rick") at Brookhaven National Laboratory on Long Island, NY.[64] In the ALICE/heavy-ion mode, the LHC will

---

collide the nuclei of lead atoms to yield 28 times the energy achievable at RHIC—a feat eagerly anticipated as the greatest energy leap in the history of particle physics.[65]

The proton-proton mode is the experimental endeavor that is associated with the theoretical black-hole disaster.[66] The LHC/ALICE experiment and the heavy-ion collisions of the RHIC are associated with a distinct doomsday scenario, the strangelet catastrophe.[67]

Putting a price tag on the LHC is difficult to do. Estimates vary, but the kind of raw power possessed by the LHC does not come cheap. The LHC machine itself, including the accelerator rings and detector arrays—what you might call the hardware—represents a cost that appears to be a minimum of about $10 billion USD.[68] The cost of the full LHC program, which includes operating the experiment and analyzing the data, is much greater.[69] If one added the cost of repurposed infrastructure from the LEP program, which includes the tunnel inhabited by the LHC, the cost would be greater still.

## C. The Payoff

So, with its voracious appetite for electricity, labor, money, and talent, what benefit does the LHC promise to deliver?

CERN supporters would say that the LHC program is part of one of the greatest, if not the single greatest, intellectual adventure in human history—the quest to understand nature at its most fundamental level. Detractors would say that the LHC, despite its enormous expense, is virtually bereft of any practical utility. Who is right? Both are.

As an effort of pure science, the LHC was not designed to have any practical utility. In this regard, particle-physics experimentation has much in common with space exploration. Historically, human space flight has spun off some valuable technologies—as has particle-physics experimentation. Proton therapy for cancer patients, for instance, is a particle-physics byproduct.[70] And then there is the Web.[71] But useful spinoffs are the exception rather than the

---

rule. The raison d'être for particle colliders is achieving a better understanding of nature—an endeavor that is intended to be its own reward.

Physics has been described as an effort to "unmask the cosmos."[72] Its importance, to its practitioners, is plain. Perhaps the most well-known expression of physicists' sentiment is a quote attributed to Ernest Rutherford: "Science is physics; everything else is postage stamp collecting."[73]

The "Holy Grail,"[74] according to physicists, would be the *theory of everything*.[75] Stephen Hawking has explained that a fully realized unified theory of nature "may not aid the survival of our species. It may not even affect our lifestyle."[76] But, "[h]umanity's deepest desire for knowledge is justification enough for continuing our quest."[77] As physicist and mathematician Henri Poincaré pointed out long ago, "The scientist does not study nature because it is useful to do so. He studies it because he takes pleasure in it and he takes pleasure in it because it is beautiful."[78]

While the LHC's start up is greeted by scientists as "unbelievably exciting,"[79] there are no guarantees of what will be found. Fermilab physicist Don Lincoln wrote that the LHC will make the universe "slightly less mysterious."[80] Precisely what knowledge will be gleaned, however, is unknown. As Lincoln explained: "We will see what we see. Interesting, fascinating, or disappointing."[81]

Even supposing the LHC is highly successful and provides a significant step on the road to a final theory, the LHC will, nevertheless, not get scientists all the way there. According to CERN physicist Michelangelo Mangano, "[w]hatever new physics is observed at the LHC, its understanding will require higher statistics and higher energies."[82]

In sum, the LHC is expected to have a handsome payoff in knowledge of fundamental particle physics, but there is no actual chance of hitting the physics-theory jackpot. Whatever the outcome, the LHC will not be the final word in physics experimentation.

---

72. *See* BRIAN GREENE, THE FABRIC OF THE COSMOS 22 (Alfred Knopf 2004) ("[A]s we've continued to unmask the cosmos, we've gained the intimacy that comes only from closing in on the clarity of truth. The explorations have far to go, but to many it feels as though our species is finally reaching childhood's end.").

73. Kolbert, *supra* note 29, at 68 (quoting Ernest Rutherford).

74. GREENE, *supra* note 13, at 15.

75. *Id.*

76. STEPHEN HAWKING, A BRIEFER HISTORY OF TIME 18 (Bantam Dell 2005).

77. *Id.*

78. LINCOLN, *supra* note 48, at xii.

79. *Id.* at 43.

80. *Id.* at 25.

81. *Id.*

82. Michelangelo L. Mangano, Physics Goals for Future Hadron Accelerators, 2nd HHH Workshop, Arcidosso 36 (2005), http://care-hhh.web.cern.ch/care-hhh/lumi-05/Plenary%20 Talks%20Wednesday/Mangano-HHH05.pdf.; *see also* LINCOLN, *supra* note 48, at ix (quoting Leon Lederman, winner of the 1988 Nobel prize in physics, "research creates new knowledge, which enables the creation of new instruments, which make possible the discovery of new knowledge").

Unless, of course, it destroys the planet.

## III. FEAR IN MANY FORMS

Understanding the black-hole controversy requires understanding its context. Particle physics has faced a long series of public-relations problems—the concern over synthetic black holes being merely the most recent. Particle physicists have worked repeatedly to try to tamp down anxieties about a parade of disaster scenarios tied to physics experimentation. At the same time, particle physicists have labored to build excitement among politicians and taxpayers for a scientific pursuit that is both extremely arcane and wildly expensive. Fear abounds on all sides. Critics, of course, fear that particle-physics experiments will annihilate humanity. At the same time, particle physicists fear that humanity will annihilate their experiments.[83]

The angst felt by the physics community is not trivial. One top CERN physicist said that most of CERN's member-nation governments are "desperately waiting for the right opportunity to shut down the place."[84]

"There is nobody fighting for this to survive, to continue," he said.[85] "We have to fight ourselves—the physicists."[86]

### A. The Strange Matter on Long Island

Besides black holes, the leading disaster scenario for particle-physics experimentation is the *strangelet scenario*. According to theory, a strangelet is a tiny, stable chunk of "strange matter."[87] Undoubtedly, a "strangelet" sounds much less frightening than a black hole. In fact, it sounds something like a benevolent character from the *Pokémon* anime series. But the cute name belies the theorized danger.

The fear is that if high-energy particle collisions created a strangelet, the object would initiate a chain reaction that would convert all of Earth and everything on it into "an inert hyperdense sphere about one hundred metres across."[88]

---

Ironically, the strangelet controversy seems to have been touched off by a physicist trying to allay fears of a black-hole disaster. In 1999, when the RHIC on Long Island was getting ready to start up for the first time, concerns were voiced that the RHIC might create black holes. A *Scientific American* reader named Walter Wagner put these concerns in a letter to the editor.[89] The magazine then published the letter in the July 1999 issue, along with a response by Frank Wilczek, a Princeton physicist who was later awarded the Nobel Prize.[90] Wilczek opined that it was not credible that the RHIC might produce black holes.[91] Then, apparently as an afterthought, he went on to say, "On the other hand, there is a speculative but quite respectable possibility that subatomic chunks of a new stable form of matter called strangelets might be produced[.]"[92]

Although Wilczek concluded in his comment that an actual strangelet disaster was "not plausible,"[93] questions about the safety of RHIC proliferated in the media.

Before the end of the month,[94] the director of the Brookhaven lab asked for a report addressing concerns about the RHIC from a committee of four scientists—Wit Busza and Robert L. Jaffe of MIT, Jack Sandweiss of Yale, and the inadvertent rabble-rouser Wilczek.[95] The resulting report, dated September 28, 1999, concluded that the possibility of a strangelet disaster was "firmly excluded" on theoretical grounds.[96] Additionally, the report authors found that on the basis of empirical evidence they could "decisively rule out" the strangelet doomsday scenario.[97]

That empirical evidence is cognizable through what is called the "cosmic ray argument."[98] Cosmic rays, despite their name, are actually particles—ones that move at ultra-fast velocities through space as they are accelerated by magnetic fields all over the galaxy.[99] The RHIC defenders pointed out that

---

everything in the galaxy is constantly bombarded with huge numbers of cosmic rays.[100] The defenders reasoned that since cosmic ray collisions mimic what particle accelerators do artificially, the survival of the Moon and other astronomical objects, despite innumerable cosmic-ray collisions, constitutes evidence that accelerators must be safe.[101]

Around the same time, a separate panel of physicists wrote a paper probing the strangelet scenario. They likewise concluded that the RHIC was safe.[102] That paper, completed in August 1999 and published as a pre-print in October 1999, was authored by three theoretical physicists at CERN. They wrote, "[O]ur extremely conservative conclusion is that it is safe to run RHIC for 500 million years."[103]

After these reports were circulated, media interest in the question of risk from the RHIC died down.

It was later revealed, however, by a physicist at Cambridge University, that both the CERN paper and the Busza report contained serious conceptual mathematical errors that unfairly downplayed the risk posed by operating the RHIC.[104] Media interest, however, did not resurface.

In 2000, a physicist named Francesco Calogero carefully ventured some of his own thoughts on the matter. A theoretician with the University of Roma–La Sapienza, Calogero had previously served as Secretary General of the Pugwash Conferences on Science and Human Affairs and accepted the 1995 Nobel Peace Prize on behalf of the Conference.[105] In his paper, Calogero criticized the findings of experts on the issues of RHIC safety, pointing to bias, a lack of scientific objectivity, and an overarching preoccupation with the public-relations consequences of what is said.[106]

Specifically, Calogero wrote that the reports on RHIC safety issues occasionally gave him "the impression that they are biased towards allaying fears 'beyond reasonable doubt.'"[107] He continued:

> I am also somewhat disturbed by what I perceive to be the lack of candour in discussing these matters of many people—including several friends and colleagues with whom I have had private discussions and exchanges of

---

messages—although I do understand their motivations. Many, indeed most, of them seem to me to be more concerned with the public relations impact of what they, or others, say and write, than in making sure that the facts are presented with complete scientific objectivity.[108]

Another person who took strangelets seriously was legal scholar and U.S. federal appellate judge Richard A. Posner, who wrote about the RHIC at some length in a 2004 book, *Catastrophe: Risk and Response*.[109] On the basis of a cost-benefit analysis he performed, Posner concluded that the RHIC was likely not worth the risk.[110]

The strangelet controversy is not limited to the RHIC. When CERN operates the LHC in heavy-ion mode using its ALICE detector, the triggering of a strangelet catastrophe is arguably a possibility there as well. Thus, it should be noted that the CERN team's paper championing the RHIC's safety was not necessarily a transatlantic act of selflessness.

Whether the LHC's ALICE/heavy-ion program will put the Earth at additional risk of a strangelet conversion is an interesting question, but one that is beyond the scope of this article.

As odd as black holes and strangelets are, they are not the only disasters hypothesized by CERN's detractors. Others are even weirder.

### B. Even Worse Than Destroying the Earth

Besides black holes and strangelets, there are three other catastrophic accident scenarios advanced as reasons to shut down the LHC. They are worth mention here because they have all been cited in LHC-injunction litigation.[111] Moreover, each of these additional scenarios commands its own share of macabre fascination. They are: magnetic monopoles, a bosenova, and a vacuum transition.

The ***magnetic monopole*** sounds perhaps even more benign than the pluckily named strangelet. But, according to hypothesis, a magnetic monopole would be just as lethal to the planet.[112] The worry is that a particle accelerator might produce a tiny bit of matter that is magnetically active, but only has one magnetic pole—that is, it would contain a net magnetic charge.[113] Where ordinary magnets have both a "north" and a "south" pole, the magnetic monopole would, for example, have only a north pole.

---

108.   *Id.* at 198.

109.   RICHARD A. POSNER, CATASTROPHE: RISK AND RESPONSE (Oxford Univ. Press 2004).

110.   *Id.* at 142. Posner's economic and mathematical analysis is discussed in some detail in Part VI., Section C, *infra*.

111.   Complaint for Temporary Restraining Order, Preliminary Injunction, and Permanent Injunction at 3-10, Sancho v. U.S. Dep't of Energy, 578 F. Supp. 2d 1258 (D. Haw. 2008) (No. 00136-HG-KSC Civ. 08) [hereinafter Complaint].

112.   CERN, The Safety of the LHC, http://public.web.cern.ch/public/en/LHC/Safety-en.html (last visited Nov. 30, 2009).

113.   *See* Hangwen Guo, Magnetic Monopoles 1, http://moreo.phys.utk.edu/adriana/electro08/guo.pdf (last visited Nov. 30, 2009).

Once produced, the fear is that a magnetic monopole would cause the protons in normal atoms to decay, initiating a runaway process that would convert and destroy the ordinary matter making up the Earth[114]—eating it up, in the words of a CERN theorist, "Pac-Man style."[115]

The argument that we have nothing to worry about from magnetic monopoles is the same as that for strangelets: If it were possible for the LHC to produce such dangerous particles, cosmic rays hitting the Earth's atmosphere would have produced them long ago.[116]

The ***bosenova*** (pronounced "BOE-suh-NO-vah") (or "bose-supernova") scenario predicts a lesser harm than destruction of the entire planet. Instead, the hypothesized harm would be like a small version of an exploding star, destroying only a piece of Switzerland and France.[117]

The alleged danger in this scenario is not the particle collisions themselves but the LHC's magnet refrigeration system and, in particular, its superfluid liquid-helium coolant.[118] It is this coolant which would hypothetically blow up supernova-style.[119] The worry stems from the fact that the ultra-cold helium is a kind of Bose-Einstein condensate,[120] a form of matter with exotic properties, which, when hit with powerful magnetic waves, can be made to explode.[121]

The bosenova disaster scenario does not have the pedigree of the other disaster scenarios; it seems to be almost entirely the work of Alan Gillis, a blogger who focuses on science issues.[122] In a paper posted in September 2008, two theoretical physicists at CERN, Malcolm Fairbairn and Bob McElrath, wrote that a bosenova cannot happen at the LHC, in part because rubidium, the

---

one substance in which a bosenova has been observed, is too different from helium.[123]

The ***vacuum transition*** scenario, also called the "vacuum bubble" or "space transition" scenario, has the unique distinction of portending something even worse than the annihilation of Earth. Specifically, a vacuum transition would destroy the entire universe.[124]

The possibility of a vacuum transition was suggested in 1984 by Piet Hut of Princeton's Institute for Advanced Study.[125] How could such a thing happen? Physicists consider empty space—"the vacuum"—to not just be a void, but to actually be *something*.[126] With the vacuum being a thing, as opposed to the absence of all things, the concern is that it might not be very stable.[127] Perhaps the vacuum's stability is, for instance, like a house of cards, stable until some movement shifts just one card, at which point the whole structure comes apart. The worry with regard to the LHC is that one of the accelerator's high-energy particle collisions might shift the fabric of space and time into a more stable state, a "vacuum bubble."[128] The universe would not disappear altogether in such an event, but it would cease to survive as we know it, and humanity would pop out of existence.[129]

CERN's argument against the possibility of a vacuum transition is, once again, the same as its argument concerning strangelets and monopoles: natural collisions of cosmic rays in our atmosphere would have already triggered such a disaster if it were possible for LHC collisions to do so.[130]

## C. The Quiet on the Texas Prairie

Despite all the doomsday scenarios described above, there is essentially only one kind of catastrophe that all particle physicists take seriously. It is a disaster of a different sort, where the only death is a figurative one: project cancellation.

No aborted experiment has ever left a deeper scar on the physics community than the United States Congress's abandonment of the Superconducting Supercollider ("SSC").[131]

---

123.   Fairbairn & McElrath, *supra* note 118, at 2–3.

124.   Piet Hut & Martin J. Rees, *How Stable Is Our Vacuum?*, 303 NATURE 508, 508 (1983) ("[The currently prevailing vacuum state] might suddenly disappear if a bubble of real vacuum formed which was large enough for the bulk energy gain . . . to exceed the surface energy density in walls.").

125.   *See* Piet Hut, *Is It Safe to Disturb the Vacuum?*, 418 NUCLEAR PHYSICS A 301 (1984).

126.   *See, e.g.,* BUSZA ET AL., *supra* note 87, at 2.

127.   Hut & Rees, *supra* note 124, at 508.

128.   *See* Hut, *supra* note 125, at 301–02; Hut & Rees, *supra* note 124, at 508.

129.   *See* Hut, *supra* note 125, at 301–02; Hut & Rees, *supra* note 124, at 508.

130.   *See* Hut, *supra* note 125, at 306–08.

131.   *See* Jeffery Mervis, *Scientists Are Long Gone, but Bitter Memories Remain*, SCIENCE, Oct. 3, 2003, at 40.

Construction on the SSC began in the early 1990s near Waxahachie, Texas, just south of Dallas.[132]

The SSC would have dwarfed the LHC.[133] A 52-mile oval tunnel—about the size of the Washington Beltway—would have accelerated two streams of protons in opposite directions at energies of 20 TeV (20 trillion electron volts) each, resulting in a collision energy of 40 TeV.[134] The SSC would have had the raw power to virtually guarantee big discoveries, including the Higgs boson.[135] But in October 1993 Congress cancelled the project, writing off the $2 billion it had already spent.[136]

The former SSC site is now an odd sort of historical monument on the Texas scrubland.[137] The large buildings left over from the project have found employment in a series of odd jobs. At one point, the site served as a warehouse for the Ellis County government, storing things like styrofoam cups.[138] In 1999, the magnet building was used as the setting for a supercomputer that commanded a robot army in a movie called *Universal Solider II*.[139] Meanwhile, deep underground, some 18 miles of tunnels have presumably flooded with groundwater.[140] In the future, the site may be used as a remote data storage center.[141]

The SSC episode left particle physicists with "bitter memories and sharp recriminations."[142] The cancellation "continues to cast a pall over their field."[143] Even a decade afterward, one particle physicist said graduate students were "staying away in droves," while established researchers were migrating to other subject areas.[144]

One member of the SSC community remarked poignantly on the suffering borne by SSC-project scientists:

> These people and their families suffered both financially and emotionally. Not only did they have to deal with rather suddenly losing their jobs, they also had to deal with the disappearance of most of the jobs in their field, as

---

universities cut back on their particle physics hiring . . . . The toll on our friends demoralized many in the field.[145]

The SSC's backers saw several factors that led to the collider's demise.[146] At least part of the blame has been placed on the fact that the physicists did not stick together. There were tensions within the physics community,[147] and a rivalry was kindled with Fermilab.[148] Moreover, some non-particle physicists spoke out against the SSC with the hope of getting some of the liberated funds if the project was cancelled.[149] As it happened, they did not.[150]

The lesson of the SSC was clear. To keep projects safe from cancellation in the future, physicists would have to pull together and work to stay one step ahead of political threats—lest particle physics be plunged into darkness.

## IV. THE DARK MENACE

The scientific issues at the heart of the black-hole question are exceedingly complicated. Particle physicists as much as acknowledge that only particle physicists have the training to understand the subject matter.[151] But it is certainly possible to sketch an outline of the controversy that is readily understandable. The scientific aspects are quite interesting—as are the human aspects.

One clear theme in the story is the pitting of insiders against outsiders. All particle physicists seem to be united in contending that there is no black-hole danger. The critics come from outside that circle, the most prominent of which are an astrophysicist, a chaos theoretician, and a mathematics professor.[152]

The other most salient theme that emerges is the demands of time. Pro-safety arguments do not evolve independently out of disinterested academic discourse; rather, such arguments are commissioned in response to media attention and published to meet accelerator-program schedules. Moreover, there is a repeating pattern of retreat and refortification in arguing the case for the safety of particle colliders. Arguments that are initially offered as unsusceptible to doubt are quietly abandoned when weaknesses are exposed. The new arguments are offered with the same sense of resolute conviction.

Understanding the controversy, however, must begin with trying to understand the thing at the center of it all—the most terrifying object humanity has ever comprehended: the black hole.

---

145.   Spencer Klein, *Letter to the Editor: The Human Cost of the SSC*, SCIENCE, Dec. 12, 2003, at 1893.

146.   *See* Mervis & Seife, *supra* note 132, at 39.

147.   *See id.*

148.   *See id.*

149.   *Id.*

150.   *Id.*

151.   Ellis, *supra* note 83 (beginning at 3 min.).

152.   *See* discussion *infra* Parts IV.D., E., H.

### A.   Behind the Event Horizon

The idea of an object exerting such enormous gravitational force that light could not escape was proposed as far back as 1783, when such a thing was referred to as a "dark star."[153] The term "black hole" was coined in the modern era by Princeton and University of Texas at Austin physicist John A. Wheeler.[154] As now understood by modern physics, a black hole is a spherical-shaped region of space within which the force of gravity is so strong that nothing—neither matter, nor light, nor information—can escape.[155] The overwhelming force of gravity causes all the matter inside a black hole to be squeezed down to an infinitely dense point at the dead center of the black hole, the "singularity."[156]

An object of any mass, if it is made small and dense enough, can form a black hole. The nine-pound, four-volume set of *Blackstone's Commentaries on the Laws of England* would form a black hole if reduced to about 10 septillionths the size of a grain of sand.[157]

Black holes have unique effects on the matter that comes their way. If you were to jump into a black hole feet first, your feet would feel the force of gravity more strongly since your feet would be closer to the singularity than your head.[158] These differential effects of gravity at varying distances, called "tidal forces," would force your feet to fall faster than your head.[159] Eventually, the tidal forces would become so great that you would be stretched out like spaghetti, your body ripped vertically into shreds.[160]

---

153.   *See* John Michell, *On the Means of Discovering the Distance, Magnitude, &c. of the Fixed Stars, in Consequence of the Diminutions of the Velocity of Their Light, in Case Such a Diminution Should be Found to Take Place in Any of Them, and Such Other Data Should be Procured From Observations, as Would be Farther Necessary for that Purpose*, 74 PHIL. PROC. OF THE ROYAL SOC'Y 35 (1784); *see also* LEONARD SUSSKIND, THE BLACK HOLE WARS 24-49 (Little Brown & Co. 2008).

154.   *See* KENNETH W. FORD, THE QUANTUM WORLD: QUANTUM PHYSICS FOR EVERYONE 233 (Harvard Univ. Press 2004); Dennis Overbye, *Physicist Who Coined the Term 'Black Hole,' Is Dead at 96*, N.Y. TIMES, April 14, 2008, at B7, *available at* http://www.nytimes.com/2008/ 04/14/science/14wheeler.html.

155.   *But see infra* Part IV.D.

156.   *See* Norton, *supra* note 14.

157.   You can calculate the Schwarzschild radius, i.e., the black hole's radius from the singularity to the event horizon, for any object for which you know the mass. Where $M$ is the mass of the object, $R_{Sch}$ is the Schwarzschild radius, $c$ is the speed of light, and $G$ is Newton's gravitational constant, the radius is found with the equation: $R_{Sch} = (2MG) / c^2$. *See* Mario Rabinowitz, *Black Hole Paradoxes*, *in* TRENDS IN BLACK HOLE RESEARCH 6 (Paul V. Kreitler ed., 2006). Here is the equation with the constants filled in, where $M$ is expressed in kilograms and $R_{Sch}$ is in meters: $R_{Sch} = (2M(6.67 \times 10^{-11})) / (3.00 \times 10^8)^2$. *See* Jearl Walker, FUNDAMENTALS OF PHYSICS A-3 (8th ed., 2008) (listing values for physical constants).

158.   GREENE, *supra* note 13, at 80.

159.   *Id.*

160.   *Id.*; HAWKING, *supra* note 76, at 81.

The idea of black holes is so horrifying that, in the beginning, physicists were skeptical that black holes might actually exist outside the imaginations of theorists.[161] Einstein famously refused to believe in their existence, even though it was his theory of special relativity that predicted them.[162]

But black holes do exist. Astronomers have now found black holes all over the universe.[163] Of course, the evidence is indirect. Because they do not allow light to escape, black holes cannot be seen directly.[164] What astronomers do see, however, is the intense light and x-rays generated as a black hole shears off the outer layers of a nearby star, accelerating the gas to nearly the speed of light while squeezing it into the gyre.[165]

## B. For the Foreseeable Future

In 1999, when questions floated in the media about accelerator-produced black holes, physicists issued an assurance that no particle collider in the foreseeable future would have enough power to accomplish such a feat.[166]

That conclusion came out of analysis performed by the authors of the Busza report, which was done in anticipation of the commencement of RHIC operations.[167] The report did a rough analysis of the particle collisions that would occur at RHIC and the gravitational effects that might result.[168] The Busza team found that the forces created by the RHIC were orders of magnitude too small to possibly create a black hole.[169] The differentials were so large, the authors said it would be "pointless to attempt refinements" of their admittedly "rough" estimating.[170] For the "foreseeable future," they wrote, an accelerator with enough power to be even remotely dangerous was "a pipe dream."[171]

Even Francesco Calogero, who expressed so much concern with regard to strangelets, was not similarly aroused with regard to black holes. Those concerns, he said, "can be allayed by simple, hence quite reliable, order of magnitude calculations, which definitely exclude any such possibility."[172]

---

But the Busza team, Calogero, and the rest were making an assumption in their calculations, an assumption that just about anyone would think is entirely reasonable. They assumed that we live in a four-dimensional world—with three dimensions of space (length, width, and height) and one dimension of time.

No one would have expected what happened next. In a plot twist worthy of *The Twilight Zone*, a problem arrived straight out of the fifth dimension.

Literally.

### C. A Problem of Many Dimensions

In 2001, a new theory concerning black holes emerged. Steven B. Giddings, a physicist from the University of California, Santa Barbara, wrote a paper with the rather provocative title, "High energy colliders as black hole factories: The end of short distance physics."[173] The paper suggested that if space had extra hidden dimensions—beyond the familiar four—then the power to make black holes might well be within grasp.[174] In particular, Giddings suggested that the LHC, when it comes online, might be able to produce black holes at the rate of one every second.[175]

Around the same time, Savas Dimopoulos of Stanford and Greg Landsberg of Brown made similar predictions in a paper called "Black Holes at the Large Hadron Collider."[176]

The idea that there might be more than three spatial dimensions may sound outlandish, but it is one of the most salient features of string theory—a burgeoning field within theoretical physics that was, in the early 2000s, gaining popularity among lay audiences. When Giddings's article came out, Columbia University physicist Brian Greene had recently published his popular, bestselling book on string theory, *The Elegant Universe*,[177] and filming was underway for a companion three-part television miniseries for *NOVA*, PBS's popular science television series.[178]

CERN had a problem. The black-hole argument used to provide cover for the RHIC was seriously undermined. A new safety rationale was needed.

---

173.    Steven B. Giddings & Scott Thomas, *High Energy Colliders as Black Hole Factories: The End of Short Distance Physics*, 65 PHYSICAL REV. D 056010, at 1 (2002); *see also* Steven B. Giddings, Some Background on the Present Work, http://www.physics.ucsb.edu/~giddings/background.html (last visited Nov. 30, 2009) (Giddings own characterization of the paper) [hereinafter Giddings's website].

174.    Giddings & Thomas, *supra* note 173, at 1.

175.    Steven B. Giddings, *Black Hole Production in TeV-scale Gravity, and the Future of High Energy Physics* ARXIV:HEP-PH/0110127v3, Nov. 1, 2001, http://arxiv.org/abs/hep-ph/0110127v3.

176.    Savas Dimopoulos & Greg Landsberg, *Black Holes at the Large Hadron Collider*, 87 PHYSICAL REV. LETTERS 161602, at A1 (2001).

177.    GREENE, *supra* note 13.

178.    *See The Elegant Universe Behind the Scenes* (PBS television broadcast Oct. 28, Nov. 4, 2003), *available at* http://www.pbs.org/wgbh/nova/elegant/maki-nf.html (last visited Nov. 30, 2009).

At some point prior to June 2002, CERN's director-general instructed an ostensibly "independent" panel of scientists[179]—though one of the six was a CERN physicist[180]—to conduct a comprehensive study regarding the safety of heavy-ion collisions at the LHC.[181]

The panel found a safety rationale that promised to cause worries about the LHC to evaporate into thin air.

### D. *Black Holes that Radiate Charm*

CERN published its safety study in 2003.[182] The study acknowledged that in the wake of advances in theory suggesting extra dimensions of space, there was a need for a "new examination of potential hazards."[183] Embarking on that examination, the report conceded that, under the new theory, black holes "*will* be produced."[184] Nonetheless, the study reported that LHC-produced black holes could not be dangerous because they would rapidly evaporate.[185] Thus, the report concluded, "black hole production does not present a conceivable risk at the LHC."[186]

Black holes evaporating? According to the classical account of black holes, evaporation is impossible—a black hole ingests everything and allows nothing out.[187] But in 1974, a young physicist at Cambridge University upended that notion. At the time, he was not well known. But he is now. The man's name is Stephen Hawking.[188]

In the work that established him as a giant in theoretical physics, Hawking argued that because of certain effects of quantum mechanics, black holes were not—and could not be—entirely black.[189] Hawking showed, mathematically,

---

that black holes must emit some form of radiation.[190] Moreover, since energy and mass are two sides of the same coin, as Einstein showed in the famous equation E=mc$^2$, then a black hole emitting radiation is a black hole that is losing mass.[191] Moreover, black holes have a temperature that varies inversely with their mass.[192] Big black holes would be extremely cold and would radiate extremely slowly.[193] Conversely, the smaller the black hole, the hotter it would be and the faster it would radiate.[194] As black holes lose mass and get ever hotter, they would start throwing off particles.[195] It is even figured that black holes that are hot enough could throw off big particles such as heavy quarks—meaning black holes would literally radiate "beauty" and "charm" particles.[196]

Given enough time, Hawking said, a black hole will heat up and evaporate right out of existence.[197]

Hawking's work has been hailed as "a brilliant tour de force" and even "the beginning of a great scientific revolution."[198]

The black-hole radiation predicted by Hawking remains theoretical—it has never been observed or experimentally confirmed.[199] Nonetheless, so-called "Hawking radiation" has reached the status of scientific orthodoxy. Few have risen to challenge it.

One of those who did was Adam D. Helfer, a mathematician at the University of Missouri.[200] In 2003, Helfer acknowledged in a paper that the prediction of black-hole radiation "is often considered one of the most secure" in its subfield of physics.[201] Nonetheless, Helfer offered his view that the prediction is based on "dubious assumptions" with unresolved difficulties.[202] "The possibility that non-radiating 'mini' black holes exist should be taken seriously," Helfer wrote.[203]

---

But few did take it seriously.

"Every so often, a physics paper will appear claiming that black holes don't evaporate," wrote Leonard Susskind, an elite physicist at Stanford. "Such papers quickly disappear into the infinite junk heap of fringe ideas."[204]

Besides, Susskind noted, black-hole radiation had been "proved" by physicist William Unruh at the University of British Columbia.[205]

Unruh's role in establishing the orthodoxy of black-hole radiation made it ironic that, after Helfer's effort, Unruh himself wrote a paper theorizing that black holes might not evaporate.[206] In 2004, Unruh, along with co-author Ralf Schützhold of the Technische Universität Dresden, concluded that "whether real black holes emit Hawking radiation remains an open question."[207]

The debate as to whether black-hole evaporation is real suddenly went from the fringe to the mainstream.

In their discussions of black-hole evaporation, neither Helfer, nor Unruh and Schützhold mentioned the safety of the LHC. But the implication for the LHC debate was clear. Walter Wagner—the same Scientific American reader who stirred the pot on strangelets and the RHIC—seized on the Unruh paper as undermining CERN's assurances of safety.[208]

Interviewed by the *New York Times* about the use of his work by LHC critics, Unruh said that the critics had missed his point.[209] For black holes to not evaporate, Unruh said, physics "would really, really have to be weird."[210]

But "weird" is not "impossible." The damage was done. While the particle-physics community might have been able to brush off most papers questioning black-hole evaporation, the paper coming from Unruh and Schützhold raised questions about the stability of LHC-produced black holes that could not be ignored.[211] That meant that to have a convincing safety rationale, CERN would have to dig deeper.

In the meantime, a new personality appeared on the side of the critics.

---

## E.  Doom, Gloom, and Smiley Faces

No one within the particle-physics community proper ventured the explicit opinion that the LHC might be unsafe. But one scientist with considerable, if unconventional credentials did sound an alarm: Otto E. Rössler of the University of Tübingen.[212] One CERN scientist later fingered Rössler as the driving force behind efforts in Switzerland and Germany to oppose the LHC.[213]

Originally trained in immunology,[214] Rössler eventually became an acknowledged pioneer in chaos theory.[215] Over the course of his career, Rössler has authored more than 300 scientific papers, and he has held professorial appointments in mathematics, chemistry, theoretical biology, theoretical biochemistry, nonlinear studies, chemical engineering, and, most relevant to this context, theoretical physics.[216] Reading about him, you begin to wonder if he wasn't the inspiration for the fictional Dr. Ian Malcolm, the eccentric chaos theorist who predicts disaster in the dinosaur thriller *Jurassic Park.*

Some measure of the man's want of conformity is found in the typography of his seminal paper on LHC safety issues: It contains three smiley-face icons.[217]

If black holes do not evaporate, as some suggested, that left open the question of how quickly they might grow—and it was this question that interested Rössler.

One might imagine that a black hole, if it does not disappear instantly, would proceed immediately to vacuum up everything nearby, quickly snowballing in size so that within a few minutes it would have consumed the entire planet. But, in fact, the growth of an accelerator-produced black hole wouldn't happen nearly so fast.[218] The most obvious thing that would hold back the growth of synthetic black holes is their initial size. Such black holes would be not just tiny, but absurdly tiny—so small, in fact, that the interior of an atom would be vast in comparison.[219]

Although counterintuitive to our everyday experience, an atom—and the tangible matter it makes up—is almost entirely empty space.[220] Electrons appear to have no size at all, and the nucleus in the middle fills about one

---

trillionth of the volume of an atom.[221] Picture a single atom as a giant sphere about two miles across, big enough to put a medium-sized airport inside.[222] The nucleus of such a magnified atom would be the size of a ball placed in the middle of all that emptiness.[223] For an element with a large nucleus, such as uranium, picture a basketball.[224] For an element with a small nucleus, such as hydrogen, picture a golf ball.[225] This illustrates how alone the nucleus is within the void of empty space in the atom's interior. Yet, such a picture does not fully explain the emptiness of an atom, because the nucleus itself is not solid; instead, it is composed of a swarm of impossibly tiny quarks and gluons that whiz about.[226] A quark is at least 1,000 times smaller than a proton—that is, if it has any size at all.[227]

Thus, a subatomic-sized black hole, even traveling through solid rock, would sail virtually unimpeded through the great voids of subatomic space, rarely encountering anything it could eat. As small as it would be, its gravity would be far too weak to attract anything. Instead, to grow, the little black hole would need to run directly into a quark, an electron, or some other elementary particle of exceedingly miniscule size.

Bearing this in mind, Rössler ran some numbers to calculate how long it would likely take black holes to grow to the point where they would be a threat. According to his calculations, LHC-produced black holes might grow fast enough that the world might end slightly more than five years after the LHC's first full-energy collisions.[228] This conclusion of Rössler's is certainly creepy. It means that people nervous about the black-hole scenario will not be able to breathe a sigh of relief even after the LHC goes to full power. The persistence of solid ground underfoot will not, unfortunately, do anything to exclude the possibility of cataclysm.[229]

One of the most interesting aspects of Rössler's paper is an espoused self-awareness of how his ideas were likely to be received. Apparently referring to himself and Helfer, the mathematician whose work he cites, Rössler said that because the LHC-risk argument "comes from a group almost devoid of credentials in the field . . . it could take years until the idea is given the benefit of the doubt."[230]

---

Consistent with prophecy, Rössler's work was in due course rejected by two major science journals.[251]

CERN scientists never responded formally to Rössler's papers, and they chose not to cite him in later papers on LHC safety.[232] Mindful of the public-relations implications of responding to Rössler directly, CERN arranged for a physicist at the Max Planck Institut für Graviationsphysik in Germany, Hermann Nicolai, to provide comments on Rössler's argument.[233] Nicolai dismissed Rössler's argument as invalid and internally inconsistent.[234] The comments of Nicolai and Nicolai's colleague, Domenico Giulini, were put into two informal documents, which then CERN made available on the web.[235] Perhaps wanting to keep Rössler's fingerprints as far from the LHC science program as possible, CERN did not post the comments in the science or safety sections of its website; instead, the comments were posted on a website devoted to showing CERN's commitment to environmental friendliness.[236]

### F. Eight White Dwarfs and Two Heroes

Even if Rössler's writings were not taken seriously, Unruh's recent writings still weakened the original safety argument of the 2003 LHC safety study. In 2007, with the LHC getting closer to completion, media and citizen inquiries into the LHC's safety led CERN management to set up the LHC Safety Assessment Group, known as "LSAG."[237] Unlike the 2003 panel, the LSAG was never presented as being independent. All five members of the group were physicists from CERN's Theory Division.[238]

---

231.   *See id.* at 1.

232.   *See* LSAG, *supra* note 179 (Rössler's work not listed in bibliography); Giddings & Mangano, *supra* note 211, at 27 (listing acknowledgements, but not citing Rössler); Peter Braun-Munzinger, Matteo Cavalli-Sforza, Gerard 't Hooft, Bryan Webber, & Fabi Zwirner ("CERN Scientific Policy Committee"), CERN, SPC Report on LSAG Documents 1 (2008), http://www.lhcfacts.org/?cat=99 (follow "The conclusion of the CERN SPC committee vetting the report" hyperlink) (no citation to Rössler).

233.   Ellis, *supra* note 83 (beginning at 108–109 min.).

234.   *Id.* (beginning at 108–109 min.).

235.   *See* Domenico Giulini and Hermann Nicolai, CERN, On the Arguments of O.E. Rössler, http://environmental-impact.web.cern.ch/environmental-impact/Objects/LHCSafety/NicolaiFurtherComment-en.pdf (last visited Nov. 30, 2009); Hermann Nicolai, CERN, Comments from Prof. Dr. Hermann Nicolai, Director, Max Planck-Institut für Gravitationsphysik (Albert-Einstein-Institut) Potsdam, Germany on speculations raised by Professor Otto Roessler about the production of Black Holes at the LHC, http://environmental-impact.web.cern.ch/environmental-impact/Objects/LHCSafety/NicolaiComment-en.pdf (last visited Nov. 30, 2009).

236.   *See* Giulini & Nicolai, *supra* note 235; Nicolai, *supra* note 235.

237.   Ellis, *supra* note 83 (beginning at 6 min.).

238.   *See* LSAG, *supra* note 179, at cover page. Note that "v2" of the report, dated September 18, 2008, removed reference to Igor Tkachev's CERN affiliation. *See* LSAG REVIEW OF THE SAFETY OF LHC COLLISIONS ARXIV:HEP-PH:0806.3414v2, CERN at cover page (2008),

In the later half of 2007, LSAG member Michelangelo L. Mangano began collaborating with Steven B. Giddings of the University of California, Santa Barbara on an in-depth theoretical treatment of LHC safety issues.[239] Giddings explained that he was motivated to work on the project after his "Black Hole Factories" paper stirred up so many questions.[240] CERN's Mangano then volunteered to join in on the project.[241]

Giddings and Mangano's work does not report Giddings's affiliation with CERN.[242] Indeed, in 2007 and 2008, when he was actually working on the paper, Giddings was not officially affiliated with CERN. But in 2006, CERN granted Giddings's application for a visiting position as a scientific associate, a position Giddings then held in 2009.[243]

As they are central figures in the black-hole case, it is interesting to get a little of the flavor of Giddings and Mangano. Both seem a good deal more hip than the stereotypical theoretical physicist, albeit in quite different ways.

Steven Giddings enjoys escaping his beach-front college campus to head up to snow-covered mountain tops; indeed he has climbed several mountains in the vicinity of CERN's base near Geneva.[244] The avid alpinist was even profiled in a *New York Times* article accompanied by a picture of him scaling a shear wall of ice.[245] In warmer locales, he has been known for his propensity to wear shorts.[246]

Michelangelo Mangano, for his part, is the picture of approachability. Most particle physicists are notoriously sloppy dressers,[247] but Mangano knows how to put on a smart suit and stylish tie and look good.[248] Speaking softly in a delightfully bouncing accent from his native Italy, Mangano is warm and relatable. It makes sense that one of Mangano's jobs at CERN is an outreach program that connects scientists with high-school physics teachers.[249] In point

---

http://arxiv.org/abs/0806.3414v2 [hereinafter LSAG Version 2]. *But cf.* Gary Felder and Igor Tkachev, *LATTICEEASY: A Program for Lattice Simulations of Scalar Fields in an Expanding Universe*, 178 COMPUTER PHYSICS COMMS. 929, 929 (2008) (listing Tkachev's institution as CERN Theory Division).

    239.    *See* Giddings's website, *supra* note 173.

    240.    *Id.*

    241.    *Id.*

    242.    *See* Giddings & Mangano, *supra* note 211 (listing only Giddings's affiliation with the University of California, Santa Barbara).

    243.    *See* Giddings's website, *supra* note 173.

    244.    *See* Steve Giddings, Climbing Highlights, http://members.cox.net/sbgiddings/climbres.html (last visited Nov. 30, 2009) (listing ascents).

    245.    George Johnson, *A Passion for Physical Realms, Minute and Massive*, N.Y. TIMES, Feb. 20, 2001, at F5.

    246.    *See* SUSSKIND, *supra* note 153, at 234.

    247.    *See, e.g.,* SHARON TRAWEEK, BEAMTIMES AND LIFETIMES 25 (Harvard Univ. Press 1992).

    248.    *See* Mangano, *supra* note 84.

    249.    *See* Michelangelo L. Mangano, http://mlm.home.cern.ch/mlm/ (last visited Nov. 30, 2009) (listing lectures he delivered in the "'High School Teachers' programme," with links to

of fact, no one could look less like a mad scientist hell-bent on building a doomsday machine than Mangano. When Oxford University's Future of Humanity Institute hosted a conference on global risks, it was Mangano who went to argue that the LHC wasn't one.[250]

After several months of work, Giddings and Mangano completed a very long and very detailed paper titled "Astrophysical Implications of Hypothetical Stable TeV-Scale Black Holes."[251] They ultimately concluded that we have nothing to worry about from the LHC.[252]

Giddings and Mangano began their work to exclude disaster scenarios by calculating how long it would take a stable microscopic black hole to grow.[253] According to their analysis, the answer depends on how many hidden dimensions there are in the universe.[254] Giddings and Mangano figured that if there are more than eight dimensions, it would take many billions of years for any black hole to accrete enough matter to be a problem.[255] For seven dimensions, they estimated the soonest the end of the Earth could come is six billion years.[256] And since the Sun is set to bloat and likely engulf the Earth within the next six billion years,[257] there would seem to be little cause for concern in a seven-dimensional universe.

But once the calculations are scaled down to a universe with six dimensions, there is a problem.[258] In a 5-D or 6-D reality, though the accretion times are still extremely long by human standards, they are much shorter than the life expectancy of the Sun. Therefore, according to Giddings and Mangano, the accretion times are "too short to provide comfortable constraints."[259]

For the 5-D scenario, Giddings and Mangano calculated a lower bound, a minimum time that would elapse before the Earth is overcome, of 300,000 years.[260] It should be noted that the 5-D scenario is certainly not implausible.[261]

---

the text of the lectures).

250.  *See* Programme of Lectures, University of Oxford Future of Humanity Institute, Global Catastrophic Risks Conference, http://www.global-catastrophic-risks.com/lectures.html (last visited Nov. 30, 2009) (Mangano delivered a lecture titled, "Expected and unexpected in the exploration of the fundamental laws of nature.").

251.  *See* Giddings & Mangano, *supra* note 211; Ellis, *supra* note 83 (beginning at 3 min.).

252.  *See* Giddings & Mangano, *supra* note 211, at 27.

253.  *Id*. at 7–14.

254.  *Id*. at 14.

255.  *Id*.

256.  *Id*.

257.  *See* MICHAEL SEEDS, HORIZONS 212 (Wadsworth 1991); David Appell, *A Solar Big Gulp: Yes, the Sun Will Eventually Engulf Earth—Maybe*, SCI. AM., Sept. 8, 2008, at 24, *available at* http://www.sciam.com/article.cfm?id=the-sun-will-eventually-engulf-earth-maybe. Some astronomers suggest we could widen Earth's orbit enough to get out of the way, although doing so would probably require sacrificing the Moon and pushing a large asteroid into an orbit where it would pass extremely close to the Earth over and over for about a billion years. *Id.*

258.  Giddings & Mangano, *supra* note 211, at 13.

259.  *Id*.

260.  *Id*.

In 2000, Giddings, along with renowned Harvard physicist Lisa Randall and MIT physicist Emanuel Katz, wrote a paper that provided an in-depth treatment of gravity and black holes in a five-dimensional universe, arguing that the 5-dimensional arrangement is possible.[262]

Faced with this potential gap in their safety argument, Giddings and Mangano went back to the drawing board to find empirical evidence of the LHC's safety. Specifically, Giddings and Mangano sought to bolster the cosmic-ray argument.[263]

To reiterate, the cosmic-ray argument provides that since high-energy cosmic-ray collisions have been happening in Earth's atmosphere throughout the planet's history, anything dangerous that the LHC could create would already have been produced by cosmic rays.[264]

There was, however, a problem with this argument for the purpose of allaying fears about LHC-produced black holes. Since the Earth is mostly empty space, it could well be incapable of stopping any black holes that are produced by cosmic rays.[265] Thus, cosmic-ray-produced black holes might exit Earth as fast as they were created.[266] LHC-created black holes, on the other hand, would not necessarily evacuate instantaneously.[267] Because protons in the LHC are propelled around the ring in opposite directions and collide nearly head-on, the momentum of each proton will largely cancel out that of the opposite proton, meaning that any resulting black holes could end up loitering in the vicinity.[268]

Thus, the old argument—that the Earth is still here despite all the cosmic rays, so we must be safe from the LHC—failed in the view of Giddings and Mangano to provide adequate assurances of safety. Even an analysis of regular stars, made of regular atomic matter that is mostly empty space, would not do the trick.[269] To make the argument work, Giddings and Mangano needed to find something in the heavens capable of stopping cosmic-ray-produced black holes.

Giddings and Mangano found what they were looking for in neutron stars and white dwarfs—special kinds of stars that are so dense, most of the empty space occurring in normal atomic matter is squeezed out.[270] For example,

---

because the matter in a neutron star is squeezed together so tightly, a tablespoon of it would weigh more than 10 trillion pounds.[271]

As it turned out, neutron stars were helpful but not definitive. Giddings and Mangano found that the persistence of neutron stars could not convincingly eliminate all LHC/black-hole scenarios because neutron stars' strong magnetic fields could slow down cosmic rays so much that, by the time the cosmic rays collided with particles inside the neutron star, the effective energy of the collision would be less powerful than the collisions planned for the LHC.[272]

Thus, it appeared that only white dwarfs could provide Giddings and Mangano with the empirical evidence they needed to close the gap and make the foolproof case that the LHC must be safe. But there are problems with the white dwarfs as well. First, white dwarfs are not nearly as dense as neutron stars.[273] Therefore, it cannot be safely assumed that any given white dwarf would trap a cosmic-ray-produced black hole.[274] Second, white dwarfs also have strong magnetic fields that could lower the energy of cosmic-ray collisions to the point where they are no longer comparable to collisions at the LHC.[275] Thus, not every observed white dwarf is an argument that the LHC is safe. Nonetheless, Giddings and Mangano identified eight observed white dwarf stars that appear to have the right properties of mass, magnetic-field strength, and age such that they present a form of living evidence that the LHC must be safe.[276]

Bottom line, Giddings and Mangano found "no basis for concerns" about black holes produced by the LHC and concluded that "there is no risk of any significance whatsoever from such black holes."[277]

LSAG was pleased. CERN's John Ellis, an LSAG member, commented in a presentation to other CERN scientists that Giddings and Mangano had done a "heroic" job.[278]

## G.  The Rise of Certitude

When they completed their paper in 2008, Giddings and Mangano submitted it to CERN's ad hoc LSAG group[279] but, in the meantime, did not disclose the paper to the public or to the broader scientific community.[280]

---

271.  *Id*.

272.  Giddings & Mangano, *supra* note 211 , at 23–24.

273.  *Id*. at 23 ("[Neutron stars] represent the highest known densities of matter that have not undergone gravitational collapse to a black hole.").

274.  *See id*. at 16–17.

275.  *See id*. at 19–23.

276.  *See id*. at 22–23.

277.  *Id*. at 27. This gives rise to the question, what do Giddings and Mangano consider *significant* in terms of risk? I will return to this question later in Part VI.C.

278.  Ellis, *supra* note 83 (beginning at 24 min.).

279.  *Id*. (beginning at 3 min.).

280.  *See id*. (beginning at 4 min.); Giddings & Mangano, *supra* note 211.

Relying on the Giddings and Mangano paper then in hand, LSAG proceeded to write a report issuing the conclusion that the LHC was safe.[281]

The LSAG Report claimed to follow on from the work of the 2003 LHC Safety Study Group.[282] In that sense, LSAG purported to "confirm, update and extend" the 2003 findings "in light of additional experimental results and theoretical understanding."[283]

But the LSAG Report's self-characterization was misleading: LSAG did not attempt to rely on the arguments from the 2003 report to justify the conclusion that the LHC is safe. The 2003 report had rested its case for safety on black-hole evaporation.[284] The 2008 LSAG report instead relied on the cosmic-ray argument as developed by Giddings and Mangano.[285]

Why did the LSAG Report retreat almost entirely to the cosmic-ray argument? Although the report doesn't say, it is not hard to guess. By 2008, the black-hole-evaporation argument had taken a bad beating. While most physicists seemed to continue to regard black-hole radiation as theoretically sound, the fact that Unruh himself began questioning black-hole radiation clearly made it less persuasive as the basis of the safety argument.[286]

What the LSAG Report did do, in lieu of arguing evaporation, was puff up the cosmic-ray argument well beyond the conclusions of the Giddings and Mangano paper.

For example, Giddings and Mangano conceded, "[W]e cannot guarantee that Earth is an efficient target for trapping hypothetical [cosmic-ray]-produced black holes in all scenarios."[287] LSAG, on the other hand, presented Earth's continuing existence as ruling out any danger: "Nature has already conducted the equivalent of about a hundred thousand LHC experimental programmes on Earth already—and the planet still exists."[288]

Additionally, while the safety argument of Giddings and Mangano ultimately funneled down to eight particular white dwarfs, the LSAG report was less discriminating. It declared that worries about "possible new particles" could be constrained or excluded by "the continued existence of the Earth and other astronomical bodies."[289]

---

281.  *See* LSAG, *supra* note 179, at cover page. Note that the report was later issued in a revised version on September 18, 2008, which included an addendum on strangelets and an updated bibliography. *See* LSAG Version 2, *supra* note 238.

282.  LSAG Version 2, *supra* note 238, at 1.

283.  *Id*. at 1.

284.  *See* LSAG, *supra* note 179, at 12 ("Thus we conclude that black-hole production does not present a conceivable risk at the LHC due to the *rapid decay of the black hole through thermal processes*.") (emphasis added).

285.  LSAG Version 2, *supra* note 238, at 3–4.

286.  *See supra* Part IV.D.

287.  Giddings & Mangano, *supra* note 211, at 14.

288.  LSAG Version 2, *supra* note 238, at 4.

289.  *Id*. at 5.

The LSAG Report's bottom-line assessment of LHC safety also went considerably further than Giddings and Mangano did. Giddings and Mangano concluded that there is "no risk of any significance whatsoever from such black holes"[290]—a statement that clearly admits of some possibility of disaster, albeit one that is insignificant.

The LSAG Report, on the other hand, used words purporting to extinguish *all* risk that could be imagined, saying that black holes from the LHC "present no conceivable danger."[291]

In May 2008, CERN's Scientific Policy Committee ("SPC") received the LSAG Report along with the Giddings and Mangano paper.[292] The SPC was pleased with what they saw, telling Giddings and Mangano that they had done a "fantastic job," and telling LSAG that they too had done a "fantastic job."[293]

The SPC then drafted its own document assessing the matter.[294] The SPC's document went further rhetorically than even LSAG was apparently willing to go. The SPC statement referred to LSAG's work as "proof" of the LHC's safety.[295] This proof, according to the SPC, relied "only on solid experimental facts and firmly established theory."[296]

The SPC's summary description of LSAG's work was favorable to the point of mischaracterization:

> For black holes, the LSAG report goes much beyond previous reports, on the basis of the GM paper. Replacing some highly plausible theoretical concepts of the previous reports with irrefutable observational data on cosmic rays and on astronomical bodies such as the Sun or compact stars, interpreted using firmly established theory, further layers of safety are added to the previously existing ones, excluding any possibility that the highly hypothetical production of black holes at the LHC could create a danger of whatever kind.[297]

This, of course, is wrong. The Giddings and Mangano paper never purported to exclude risk based on the existence of bodies such as the Sun, doing so instead on the basis of white dwarfs and neutron stars.[298] It is also dubious to conclude that "any possibility" of disaster was excluded when the Giddings and Mangano paper claimed only to exclude "risk of any significance."[299] Moreover, the characterization of the observational data as

---

"irrefutable" came from the SPC.[300] The word Giddings and Mangano used to characterize the astronomical data was "solid."[301]

Thus, having ascended through the CERN hierarchy, the likelihood that the LHC was dangerous went from *insignificant* to *inconceivable* to *impossible*.

After the SPC favorably reviewed the Giddings and Mangano article and the LSAG report, the papers were then disclosed to the public in 2008.[302] At that point, one individual took up the task of lending the Giddings and Mangano paper a critical eye.

## H.  The Polite Outsider

Rainer Plaga is an astrophysicist, formerly of the Max-Planck Institute and currently working for the German federal government as a civil servant.[303] In August 2008, Plaga completed a paper—in his capacity as an individual and not on behalf of the government—that took issue with the conclusions of Giddings and Mangano. The approach Plaga took differed significantly from that of Rössler. Plaga's paper was less alarmist, more deferential, and much more carefully presented.[304] The paper was also instantly visible to the particle-physics community because Plaga posted it online on the *arXiv*, a hotbed for developing scholarship in physics and mathematics that functions as a self-publication method for pre-prints of articles not yet published in a conventional print journal.[305]

Plaga's approach was also different in that it focused specifically on the latest definitive treatment of LHC safety: the Giddings and Mangano paper, which he referred to respectfully, complimenting it as "excellent."[306] Plaga's paper, published only seven weeks after Giddings and Mangano's paper, presented two independent arguments against the conclusion that the LHC is safe.[307] Optimistically, Plaga did not predict that the LHC would destroy the world; instead, he argued that Giddings and Mangano's paper did not exclude all possibilities of disaster—that is, that it left "a residual catastrophic risk" on the table.[308]

---

300.    CERN Scientific Policy Committee, *supra* note 232, at 4.

301.    Giddings and Mangano, *supra* note 211, at 1.

302.    *See* Ellis, *supra* note 83 (beginning at 11–12 min.); Giddings and Mangano, *supra* note 211, at 1.

303.    *See* Rainer Plaga, *Astrophysics: Rays from the Dark*, 453 NATURE 48, 49 (2008) ("Rainer Plaga is in the Department for New Technologies and Scientific Foundations, Federal Office for Information Security."); Max-Planck-Institut Für Physik Werner-Heisenberg-Institut Jahresbericht Annual Report 2000, *available at* http://www.mppmu.mpg.de/english/jbericht-00.pdf (last visited Nov. 30, 2009).

304.    Rainer Plaga, *On the Potential Catastrophic Risk from Metastable Quantum-black Holes Produced at Particle Colliders*, ARXIV:GEN-PH/0808.1415v2 (Sept. 26, 2008), *available at* http://arxiv.org/abs/0808.1415v2. Note that the first version of the paper was published on August 10, 2008, and can be found at http://arxiv.org/abs/0808.1415v1. Citations herein are to the second version unless noted.

305.    *See* arXiv.org, Cornell University, http://arxiv.org/ ("arXiv is an e-print service in the fields of physics, mathematics, non-linear science, computer science, quantitative biology and statistics.").

306.    Plaga, *supra* note 304, at 8.

Plaga's first argument was based on an alternative model of microscopic black holes, one previously advanced by Roberto Casadio and Benjamin Harms in 2002.[309] On this basis, Plaga concluded that growth of microscopic black holes might not be slow and that black-hole behavior that would destroy the Earth might not have the same effect on white dwarfs.[310] Thus, the eight white dwarfs pointed to by Giddings and Mangano might not be conclusive empirical evidence of the LHC's safety. Plaga then made the eye-opening suggestion that as a consequence of the alternative black-hole model, a fast-growing black hole could cause the release of an amount of energy equal to 12 megatons of TNT per second, appearing "like a major nuclear explosion in the immediate vicinity of the collider."[311]

Plaga's second argument, independent of the alternative-model argument, took issue with a mathematical assumption made by Giddings and Mangano regarding the size of the black holes.[312] Specifically, Giddings and Mangano assumed that the black hole would have a certain minimum size—three times as large as the "Planck scale."[313] Plaga suggested that while this might be a good assumption for pure research purposes, such as analyzing data collected from a particle accelerator detector, it is not an appropriate assumption for purposes of disaster-risk analysis.[314]

So, what is the significance of the Planck scale? The Planck scale defines a realm of the incredibly tiny.[315] Distances measured in terms of the Planck scale are small even by particle-physics standards.

---

Below the Planck scale, in the words of renowned physicist Brian Greene, "quantum uncertainty renders the fabric of the cosmos so twisted and distorted that the usual conceptions of space and time are no longer applicable."[316] The Planck scale "describes an unfamiliar arena of the universe in which the conventional notions of left and right, back and forth, up and down (and even of before and after) lose their meaning."[317] According to Greene:

> Physicists believe . . . the smooth portrayal of space and time is an approximation that gives way to another, more fundamental framework when considered on ultramicroscopic scales. What that framework is—what constitutes the "molecules" and "atoms" of space and time—is a question being pursued with great vigor. It has yet to be resolved.[318]

With specific reference to black holes, Plaga, citing prior work by Giddings, contended that contemporary physics is unable to reliably predict the behavior of smaller black holes; thus, one cannot logically rule out the possibility that such black holes can be produced.[319] Further, Plaga claimed that smaller black holes are not excluded by Giddings and Mangano's white-dwarf analysis.[320] Why not? The problem with smaller black holes is that they might be able to slip through white dwarfs without being trapped, Giddings and Mangano's analysis notwithstanding.[321] If that is true, then the existence of white dwarfs does not prove that cosmic-ray collisions do not produce black holes. If cosmic-ray collisions do, in fact, produce black holes, so could the LHC, and LHC-produced black holes could remain Earth-bound.[322]

In closing, Plaga offered some compromises, which he called "feasible measures for risk mitigation" that would not force a total shutdown of the LHC program.[323] He proposed that the LHC ramp up slowly, proceeding incrementally, rather than going immediately to full power.[324]

While Rössler was mostly ignored, Plaga was not. Within 19 days after Plaga posted his paper—and a little more than a week before the LHC's planned start up—Giddings and Mangano posted a paper on the *arXiv* disputing Plaga's alternative-model argument.[325]

Giddings and Mangano's reply paper was interesting in both tone and content. The tone was confrontational and decidedly icy. In terms of content,

---

316.  GREENE, *supra* note 72, at 334.
317.  GREENE, *supra* note 13, at 129.
318.  *Id.* at 335.
319.  Plaga, *supra* note 304, at 7–8.
320.  *Id.* at 8.
321.  *Id.*
322.  *Id.* at 7–8.
323.  *Id.* at 9.
324.  *Id.*
325.  *See* STEVEN B. GIDDINGS AND MICHELANGELO L. MANGANO, COMMENTS ON CLAIMED RISK FROM METASTABLE BLACK HOLES, ARXIV:HEP-PH/0808.4087V1, CERN-PH-TH-2008-184, CERN (CERN Rep. No. CERN-PH-TH/2008-184) (2008), http://arxiv.org/abs/0808.4087v1.

Giddings and Mangano responded to only some of Plaga's arguments, declining to discuss or even mention others.[326]

With regard to Plaga's argument based on the Casadio and Harms model, Giddings and Mangano first claimed that Plaga applied a formula inconsistently.[327] Next, they offered two more arguments to refute Plaga's conclusions but provided only a single sentence of explanation for each, saying they would "defer further explanation for future comment."[328] First, they argued that an approximated equation provided in their appendix can be applied to exclude the risk Plaga discussed.[329] Their second argument was that the underlying basis of Plaga's paper "appears implausible."[330]

Giddings and Mangano ended their comment by accusing Plaga of misquoting their paper and selectively quoting from the available literature.[331] The accusations are worth scrutinizing. For the misquote, Giddings and Mangano recited the correct text, but did not provide the context of Plaga's misquote or explain its significance.[332] The result was that Giddings and Mangano created the impression that the misquote was material, when in fact, it was quite insignificant.[333] It turns out that all Plaga did was drop an "or" in a sentence, which had the effect of making Giddings and Mangano's paper look, if anything, slightly more persuasive than it would have appeared otherwise.[334]

Even more interesting was the accusation that Plaga was selectively quoting from the available literature. Where Plaga cited to Unruh's 2004 paper questioning the existence of Hawking radiation,[335] Giddings and Mangano suggested he should have provided "the more recent citation to Unruh's work," which "reflects more up-to-date comments by Unruh on the support of his work for Hawking radiation."[336]

---

326. *Id.* at cover page ("In a recent note . . . it was argued that a hypothetical metastable black hole scenario could pose collider risk not excluded by our previous study. We comment on inconsistency of this proposed scenario.").

327. *Id.* at 1–2.

328. *Id.* at 2.

329. *See id.*

330. *Id.*

331. *Id.*

332. *See id.*

333. *See id.*

334. Plaga's misquote: "at each point where we encountered an uncertainty, we have replaced it by a conservative 'worst case' assumption." Plaga, *supra* note 304, at 3 n.3 (First version). The correct quote is: "at each point where we encountered an uncertainty, we have replaced it by a conservative *or* 'worst case' assumption." Giddings and Mangano, *supra* note 211, at 2 (emphasis added). The misquote made the statement more plausible because, where someone is arguing the case for safety, it is more persuasive to utilize only *conservative worst-case* scenarios rather than some scenarios that are not *worst case*, but are, at least, *conservative.* Plaga corrected the misquote in the second version of his paper. Plaga, *supra* note 304, at 3 n.3.

335. Unruh & Schützhold, *supra* note 206.

336. Giddings & Mangano, *supra* note 325, at 2.

The "more up-to-date" citation to which Giddings and Mangano referred was not a paper at all, but an abstract of a talk Unruh gave in 2007.[337] In fact, it does not repudiate Unruh's prior work.[338] It is also hardly a ringing endorsement of Hawking radiation. In favor of Giddings and Mangano's case, the abstract says, "Analog models of gravity have given us a clue that despite the shaky derivation, the effect is almost certainly right."[339] But the abstract also says, "Black Hole evaporation is one of the most puzzling features of gravity and quantum theory."[340] The abstract goes on to say that the theory's derivation by Stephen Hawking is "nonsense, in that it uses features of the theory in regimes where we know the theory is wrong."[341]

There remains one more aspect to discuss about the Giddings and Mangano rebuttal—something that is rather remarkable: Giddings and Mangano offered no response at all to Plaga's second line of argument, which took issue with Giddings and Mangano's mathematical assumptions regarding black-hole size.[342] Thus, Plaga's argument regarding black-hole size in relation to the Planck scale stands as a level of uncertainty about LHC safety to which CERN has made no response.

Plaga posted a sur-reply to Giddings and Mangano in an appendix to a second version of his paper, which corrected the misquote.[343] Plaga also insisted that he did not employ the equations inconsistently, and Plaga argued that Giddings and Mangano criticized him for making mathematical moves he never employed.[344]

The debate with Plaga ends there. Giddings and Mangano left Plaga's most recent points without a response. Perhaps Giddings and Mangano judged that, given his marginalized outsider status, engaging in further dialog with Plaga would only bolster the case that his arguments should be taken seriously.

## I.    *Warped in the Brane*

After the release of "Astrophysical Implications" and the LSAG Report, physicists piled on with concurrences that the LHC was safe.

---

The Executive Committee of the Division of Particles & Fields of the American Physical Society endorsed the LSAG Report, as did the UK Institute of Physics.[345]

Michael E. Peskin of Stanford offered a viewpoint piece on the Giddings and Mangano article, praising it for taking doomsday hypotheses "as a challenge" and using them to launch "a new and fascinating investigation."[346]

With the stated aim of assuaging "public alarm" about the LHC, Benjamin Koch, Marcus Bleicher, and Horst Stöcker of the Johann Wolfgang Goethe–Universität in Frankfurt published a paper that summarized existing safety arguments and added new ones.[347] The paper lauded the LHC and concluded that there was no logically possible black-hole evolution path that could be dangerous.[348] The paper's acknowledgement section was noteworthy for its marked duo-tone contrast: "The authors thank Michelangelo Mangano for fruitful discussions and his helpful comments. We also acknowledge discussion with Rainer Plaga."[349]

Sergio Fabi and Benjamin Harms of the University of Alabama and Roberto Casadio of the Università di Bologna published a paper from a string-theory perspective, arguing against the possibility of danger at the LHC.[350] The paper, titled "On the possibility of Catastrophic Black Hole Growth in the Warped Brane-World Scenario at the LHC," was based on their "previous study of black holes in the context of the warped brane-world scenario."[351] While the talk of "warped branes" sounds like an ad hominem attack on LHC detractors, it is not. The word "brane" is a term of art in string theory referring to a kind of cosmological structure existing in a higher-dimensional universe, and "warped" describes a geometric quality, not a psychological one.[352]

Yet there were ad hominem attacks. John Ellis of CERN referred to LHC detractors as "nuts" and insinuated that one of them, Walter Wagner, was only

---

345.  *See generally* Press Release, Division of Particles & Fields of the American Physical Society, Statement by the Executive Committee of the DPF on the Safety of Collisions at the Large Hadron Collider, http://www.aps.org/units/dpf/governance/reports/upload/lhc_saftey_statement.pdf (last visited Nov. 30, 2009); Press Release, The Institute of Physics, LHC Switch-on Fears are Completely Unfounded (Sept. 5, 2008), http://www.iop.org/Media/Press%20Releases/press_31275.html.

346.  Michael E. Peskin, *The End of the World at the Large Hadron Collider?*, 1, 14 PHYSICS 1 (2008), *available at* http://physics.aps.org/pdf/Physics.1.14.pdf.

347.  *See* Benjamin Koch et al., *Exclusion of Black Hole Disaster Scenario at the LHC*, ARXIV:HEP-PH/0807.3349v2, Sept. 27, 2008, http://arxiv.org/abs/0807.3349v2.

348.  *Id.* at 13.

349.  *Id.*

350.  *See* Roberto Casadio et al., *On the Possibility of Catastrophic Black Hole Growth in the Warped Brane-World Scenario at the LHC*, arXov:0901.2948v2, Feb. 17, 2009, http://arxiv.org/PS_cache/arxiv/pdf/0901/0901.2948v2.pdf (accepted for publication in Physical Review D on August 14, 2009).

351.  *Id.* at 1.

352.  GREENE, *supra* note 13, at 414.

pursuing a lawsuit against CERN to make money.[353] Yet Wagner was suing for an injunction, not damages.[354] Another CERN physicist referred to LHC critics as "crazy people."[355] Much more blunt was renowned University of Manchester physicist Brian Cox:

"Anyone who thinks the LHC will destroy the world," he said, "is a twat."[356]

### J.  The Best Answer

In August 2008, with the public opening of the LHC coming within weeks, John Ellis gave a talk in the CERN auditorium in which he sought "to provide the ammunition" that CERN people could use to convince others that the collider poses no danger.[357]

After reviewing the scientific arguments that the LHC is safe, Ellis explained that a question that worried him more than whether humanity was safe from the LHC, was the opposite—whether the LHC was safe from humanity.[358]

Ellis then briefed the audience on unfavorable press reports, the various lawsuits filed to stop the LHC, and a public opinion poll indicating that most people thought the LHC was not worth the risk.[359] Ellis also introduced the audience to Richard Posner, whom Ellis said he found "really worrying."[360]

Wrapping up, Ellis came to what his presentations slides labeled "The Best Answer."[361]

"So, to finish," Ellis said, "the way to stop all this argument about whether the LHC is going to destroy the planet is to get the LHC working."[362]

"Within a few weeks time, we will know that LSAG was right."[363]

Of course, in making such a statement, Ellis either showed that he misapprehends the relevant physics, which seems highly doubtful, or he revealed himself to be disingenuous. As the Giddings and Mangano report acknowledged, black-hole disaster scenarios provide for long incubation times, during which microscopic black holes may reside in the Earth unnoticed,

---

slowly gaining mass.[364] Therefore, starting the LHC would actually do nothing to prove that the LHC was safe. Of course, it might well end the debate by *fait accompli*.

At any rate, as it turned out, CERN came up short of providing this "best answer." In the ensuing weeks, CERN was unable to get the LHC working.[365]

During testing and before particle collisions began, a faulty electrical connection caused an electrical arc, puncturing the LHC's superfluid helium enclosure and releasing a large amount of helium coolant into the LHC tunnel.[366] The accident was violent enough to rip a portion of the LHC out of its anchors in the concrete floor.[367] The necessary repairs have required the LHC's first collisions to be pushed back more than full year.[368] CERN also had to downgrade its initial expectations for the machine, opting to begin operating at reduced energies in order to get started collecting at least some data.[369]

Ultimately, CERN decided that over the 2009-2010 run, the LHC would collide particles at a maximum energy of 7 TeV[370]—half the machine's designed strength of 14 TeV.[371]

In mid-October 2009, the LHC was finally cooled down to its ultra-refrigerated operating temperature,[372] and in mid-November, CERN succeeded in steering particles half-way around the LHC ring.[373] As of late November 2009, CERN's plan is to commence particle collisions before Christmas.[374]

---

In the meantime, the litigation to stop the LHC continues to simmer in the courts.

## V. PROBLEMS FOR PLAINTIFFS

Suing to stop the LHC is a unique litigation endeavor. Problems abound. The only thing that seems straight-forward is the prayer for relief. But what is the claim? In what court do you file it? And how do you get personal jurisdiction over CERN?

### *A. The Litigants' Attempts*

Multiple suits to stop the LHC have been initiated by a colorful assortment of plaintiffs. An action was brought in the Swiss courts in 2008, but it was dismissed because of CERN's treaty-based immunity from process in Switzerland.[375] A lawsuit in Germany seeking to force the German government to use its membership in CERN to prevent full-power operation has also been unsuccessful—as of early 2009, the case was still alive, as the plaintiffs had appealed to Germany's *Bundesverfassungsgericht* (Federal Constitutional Court).[376] A third action was instituted in the European Court of Human Rights ("ECHR"), seeking "interim measures" to halt the LHC's planned operation.[377] The ECHR dismissed the interim-measures request with a brief e-mail, which stated no reasons for the decision.[378]

In Hawaii, Walter Wagner, along with Luis Sancho, a self-described "leading researcher in the field of Time Theory,"[379] filed a pro se complaint in the United States District Court for the District of Hawaii.[380] The plaintiffs attempted to bridge the transoceanic divide and reach CERN by invoking the National Environmental Policy Act[381] ("NEPA").[382] Through NEPA, the plaintiffs sought to require an environmental impact statement from the U.S. government in its role of funding and participating in the LHC project.[383]

---

375.    *See* Klage betreffend Gefährdung meines Lebens, Bezirksgericht Zürich, June 3, 2008 (*Schröter v. CERN*) at 1, 3 (on file with author).

376.    *See* Verfassungsbeschwerde, Antrag auf Erlass einer einstweiligen Anordnung und Prozesskostenhilfegesuch, Bundesverfassungsgericht, Jan. 21, 2009 (*Schröter v. Bundesrepublik Deutschland*); In dem Verfassungsbeschwerdeverfahren und Verfahren auf Erlass einer einstweiligen Anordnung, Bundesverfassungsgericht, Mar. 10, 2009 (*Schröter v. Bundesrepublik Deutschland*) (on file with author).

377.    *See* Vorab Per E-mail, European Court of Human Rights, Beschwerde Nr. 41028/08, Aug. 29, 2008 (*Goritschnig v. Österreich*) (on file with author).

378.    *See id.*

379.    *See* Affidavit of Luis Sancho in Support of TRO and Preliminary Injunction at 2, Sancho v. U.S. Dep't of Energy, 578 F. Supp. 2d 1258 (D. Haw. 2008) (No. CV08-00136-HG-KSC).

380.    *See* Sancho v. U.S. Dep't of Energy, 578 F. Supp. 2d 1258, 1259 (D. Haw. 2008).

381.    National Environmental Policy Act of 1969, 42 U.S.C. §§ 4321–4347 (2006).

382.    *See Sancho*, 578 F. Supp. 2d at 1259, 1265.

While *Sancho v. Department of Energy* had all the hallmarks of a case that would not be taken seriously by the court, Judge Helen Gillmor, as indicated by her opinion, handled the case conscientiously and thoughtfully.[384] Notably, she did not disparage the merits of the plaintiffs' claim, writing, "It is clear that Plaintiffs' action reflects disagreement among scientists about the possible ramifications of the operation of the Large Hadron Collider. This extremely complex debate is of concern to more than just the physicists."[385]

Yet the outcome was still what one would tend to expect—the case was dismissed on jurisdictional grounds.[386] The federal government's funding and involvement did not provide a sufficient nexus for substantive jurisdiction under NEPA.[387] Gillmor's opinion suggested that the political process was the appropriate forum for airing what it characterized as a policy disagreement.[388] Thus, in American courts, the question remains: What cause of action could be used to force CERN to defend a suit on the merits?

## B.  Stating a Claim

If the LHC destroyed the planet, there would be no *legal* problem (though the practical problem is plain) with suing CERN for damages. Getting an injunction ahead of time, however, requires a bit more thought.

The case for an injunction would be easy, of course, if CERN explicitly threatened to intentionally destroy the Earth. Threatening an illegal act is grounds for an injunction,[389] and compacting the Earth down into a marble-sized black hole would involve a significant amount of trespass, battery, and conversion, not to mention the violation of an impressive array of regulations, ordinances, and criminal statutes. But the only action CERN has signaled it will intentionally do is turn on the LHC. Thus, to get an injunciton on the basis of a threat of illegal action, plaintiffs would have to show that CERN, or some other necessary party, would be violating a law merely by starting up the LHC.

Suppose there were a Worldwide Accelerator Safety Administration that promulgated safety standards having the force of law.[390] If the LHC violated

---

383.    *See id.* at 1262–63, 1265.

384.    *See generally id.* at 1258.

385.    *Id.* at 1269.

386.    *See id.* at 1268.

387.    *See id.*

388.    *Id.* at 1269 (quoting Metro. Edison Co. v. People Against Nuclear Energy, 460 U.S. 766, 777 (1983)).

389.    Threatening an illegal act is grounds for an injunction if the harm is irreparable and the other requirements are met. *See, e.g.*, Hurley v. Bd. of Miss. Levee Comm'rs, 23 So. 580, 581 (Miss. 1898); Smart v. Hart, 44 N.W. 514, 514 (Wis. 1890).

390.    Euan MacDonald suggested, in response to my *PrawfsBlawg* posts about the LHC, that a global administrative law regime would be well-suited to dealing with particle accelerator one of those standards, an injunction would be a slam dunk. There being no such thing, the *Sancho* plaintiffs tried using NEPA to fill the gap, but the attenuated nature of the U.S. government's involvement stretched the claim too far.[391] Besides, even if a NEPA violation were found, the statute would only require a pause in the LHC program while an environmental-impact statement was prepared.[392] Perhaps such a delay might provide time to galvanize public opposition and bring about a permanent shut down of the LHC through political means. But in such a case, the ultimate authority would not lie with the courts. From a legal standpoint, the more compelling question to consider is whether the courts, by themselves, have the inherent power to enjoin an activity solely on the basis of a perceived risk.

There are at least two common contexts in which such injunctions frequently arise: domestic violence and trade secrets. Domestic-violence injunctions, usually termed "restraining orders" or "orders of protection," are an everyday occurrence.[393] Authorized by statute, such injunctions do not require that physical injury has been previously suffered or even specifically threatened.[394] Thus, such injunctions have an independent legal existence in that they do not depend upon the violation or the explicitly threatened violation of any law. Trade-secret injunctions are also an unremarkable occurrence. Courts may issue prophylactic injunctions where there is a danger that trade secrets may be misappropriated in the future—even where it is stipulated that

---

safety issues. Posting of Euan MacDonald to Global Administrative Law, http://globaladminlaw.blogspot.com/2009/01/global-administrative-law-and-end-of.html (Jan. 20, 2009, 13:47 EST).

391. *See Sancho*, 578 F. Supp. 2d at 1266–67.

392. *See generally* 42 U.S.C. § 4332(C)(i) (requiring an environmental impact statement for "major Federal actions significantly affecting the quality of the human environment"); COUNCIL ON ENVIRONMENTAL QUALITY EXECUTIVE OFFICE OF THE PRESIDENT, A CITIZEN'S GUIDE TO THE NEPA: HAVING YOUR VOICE HEARD 2 (2007), http://ceq.hss.doe.gov/nepa/Citizens_Guide_Dec07.pdf (noting that "NEPA requires agencies to undertake an assessment of the environmental effects of their proposed actions prior to making decisions").

393. *See, e.g.*, Beth I.Z. Boland & Susan M. Finegan, *Survey of Key Developments in the SJC's Recent Approach to Domestic Violence Issues:* Jacobsen, Frizado, Kwiatkowski, *and* R.H. v. B.F., BOSTON B.J. 10, 10, Jan.–Feb. 1996 (finding that 46,265 restraining orders were issued in Massachusetts during a one-year period in 1994 and 1995); John W. Fountain & Joseph Kirby, *Stalking Victims Find Laws Are Little Help*, CHI. TRIB., Aug. 5, 1992, at D1 (reporting that 35,346 violations of orders of protection occurred during a five-month period in 1992); *Record Number of Restraining Orders Issued in S. Florida*, PALM BEACH POST, Mar. 17, 1992, at 4B (stating that, in a three-county area of Florida, 8,224 restraining orders were issued in 1991); Lisa Redmond, *When Dating Goes Bad, Teens Seek Date in Court*, LOWELL SUN 1, Aug. 17, 2006, *available at* http://www.ncdsv.org/images/When%20Dating%20Goes%20Bad_Teens%20Seek%20Date%20in%20Court.pdf (noting that there were in excess of 32,000 restraining orders issued yearly in Massachusetts in 2001, 2002, and 2003).

394. *See, e.g.*, MD. CODE ANN., FAM. LAW §§ 4-501(b)(1)(ii), 4-505, 4-506 (West 2006); W.VA. CODE ANN. §§ 48-27-101, 48-27-202, 48-27-501, 48-27-502 (Lexis 2004); WYO. STAT. ANN. §§ 35-21-102(a)(iii)(B), 35-21-104, 35-21-105 (2009).

the defendant has not used and does not intend to use certain confidential information.[395]

Outside of these specific contexts, injunctions against a mere chance of future injury are rare. The question for the black-hole case is: Do the courts have the general power to enjoin conduct simply because it is adjudged to be unreasonably unsafe—in other words, can courts enjoin *negligence*?

Issuing an injunction against negligence—as such, and without more—is a concept that is largely foreign to American jurisprudence. One commentator has remarked, "It is well settled that equity will not interfere when the apprehended injury or harm is doubtful or speculative."[396] But such a statement goes too far. Under the right circumstances, though they are quite infrequent, the courts will indeed enjoin negligence. It is, however, doubtlessly fair to say that it is generally the case that the courts will not use equity where the harm is speculative. But why?

The first reason has to do with where the uncertainty lies in the injunction case. In the most ordinary sort of injunction case, uncertainty arises with regard to the *law*. But when negligence is to be enjoined, the uncertainty arises with regard to the *facts*—specifically, future facts. For example, when a group of homeowners seeks to enjoin an airport from extending its hours of operation, the question is not whether extended operation will produce additional noise—that much is clear—the question, instead, is whether the homeowners have a legal right that will be violated by the additional landings and take-offs. Courts can easily solve legal uncertainty—that is their forte.

Resolving factual uncertainty, however, is more difficult. Moreover, the factual uncertainty that must be wrestled with when considering an injunction against negligence is more challenging than that which must be dealt with when deciding an *ex post*[397] negligence case as a matter of law. This additional level of difficulty arises because of the particular kind of factual uncertainty involved. Where factual uncertainty exists in an *ex post* negligence case, it will not regard whether or not an injury was suffered. With rare exceptions, the injury in an *ex post* negligence case is undisputed fact. But in an *ex ante*[398] negligence injunction case, the injury itself is uncertain, and injury is an indispensable element of a negligence cause of action.[399]

---

Allowing a cause of action for an unaccrued injury is troublesome because even if a court were to decide that a particular course of action is negligent—that is violative of a standard of reasonable care with regard to a plaintiff to whom the defendant owes a duty—it then remains forever uncertain as to whether a protectable interest would have been invaded. For instance, a plaintiff may drive negligently all day long without causing any damage or injury. In such a case, we would agree that no legal interest has been invaded—unless and until someone actually gets injured as a result of the negligent driving.

The second reason that negligence generally is not proscribed by injunction is that the type of situation that would present such a claim almost never arises. In reality, there is seldom an occasion to ask a court to use its equity powers to enjoin negligence. When people desire safety—negligence's antonym—regulations and legislative solutions are the habitual resort. Even people who anticipate injury do not generally know who will injure them. A driver prospectively fearful of injury from vehicular negligence, for example, would not know whom to sue. Thus, traffic policing and vehicle safety codes are the usual legal means for decreasing the chance of injury on the road.

Even in those instances where a particular defendant is identifiable, criminal and regulatory schemes almost always fill the need. Imagine having a next-door neighbor who drinks heavily at night and discharges firearms in his front yard while your kids are playing outside. Is your neighbor acting negligently? Yes. But is an injunction necessary? Hardly. A simple call to the police will take care of the problem. The police have at their disposal a myriad of ordinances and state laws to use in charging the miscreant.

Relatedly, where safety is concerned, those agitating for change ordinarily take on their task for the benefit of the general public against a multiplicity of actors. Aggrieved parents or loved-ones often seek regulatory or statutory reform that will, in their view, prevent the kind of accident that caused their suffering. Thus, safety concerns generally translate into work for lobbyists rather than courtroom lawyers. In other words, ex-ante concern often makes for agitants, but it rarely makes for litigants.

Thus, cases which can be said to establish grounds for enjoining negligence in the rarefied air of pure equity—and not within a statutory or regulatory framework—are sparse.[400] When posited, the question of enjoining negligence puts in conflict two important sets of values. On the one hand, the law has a general disinclination to injunctions. As restrictions on liberty, American law finds injunctions inherently uncomfortable. On the other hand, the courts are loathe to gamble with human life and limb. In considering these values, it follows that courts must, despite discomfort, enjoin negligence on those occasions when equity requires.

---

400.	It should be noted that the rarity of reported cases does not necessarily mean that grants of equitable relief in the trial courts are nearly so uncommon.

Indeed, the reporter volumes contain several examples where courts have done so. Courts have approved injunctions where the where a dangerous child sought to attend school,[401] and where a telephone-company employee with health issues sought to avoid the hazard of other employees smoking tobacco near her.[402] In one case, a federal district court enjoined the proposed placement of a homeless shelter next to a psychiatric hospital on the basis that it put patients at significant risk, writing, "Activities which threaten human life are cognizable and may be enjoined even before the threat matures to result in physical injury or death."[403]

A case that is closer to the facts of the black-hole cases—though one that ultimately resulted in the denial of an injunction—is *Brennan v. Gellick*.[404] Decided in 1892, it is a case of its era, one that would almost certainly not come to court today, since modern safety regulations would render it pointless. In *Brennan*, the court declined to enjoin the continued blasting of bedrock, even though the blasting that had already been done had caused considerable damage to the plaintiff's neighboring property.[405] The court identified powerful policy interests against such an injunction, figuring that "[t]o hold that blasting in the City of New York is intrinsically dangerous and unlawful would be to put an end to all public improvements."[406] But *Brennan* does not stand for the proposition that negligence cannot be enjoined. To the contrary, the court left open such a possibility using analysis quite compatible with an injunction in the black-hole case.[407] The *Brennan* court found that on the facts before it, the plaintiff actually was without "adequate cause to fear irreparable injury," therefore making an injunction unwarranted.[408] The plaintiff was invited to re-apply if sufficient evidence eventually accumulated.[409]

The right set of facts for an injunction arrived more than 50 years later in the case of *Harris Stanley Coal & Land Co. v. Chesapeake & Ohio Railway Co.*[410] The dispute arose in mountainous eastern Kentucky, with the plaintiff railroad suing for an injunction to stop the defendant coal mine from underground pillar-pulling operations, which, according to the railroad's allegations, would cause surface subsidence and landslides that would injure railroad property and operations.[411] Notably, the case arose during World War II, a time of increased demand for coal-based energy.[412]

---

401.   *See* Honig v. Doe, 484 U.S. 305, 327 (1988) (recognizing "the equitable powers of district courts [to] temporarily enjoin a dangerous disabled child from attending school.").

402.   Shimp v. New Jersey Bell Tel. Co., 368 A.2d 408 (N.J. Super. Ch. 1976) (recognizing a "common-law right to a safe working environment" and finding that "such work area was unsafe due to a preventable hazard which the court had power to enjoin."); *see also* Smith v. Western Elec. Co., 643 S.W.2d 10 (Mo. App. 1982).

403.   Seide v. Prevost, 536 F. Supp. 1121, 1133 (S.D.N.Y. 1982) (denying a request for an injunction, made by patients at a psychiatric hospital to prevent the erection of a homeless shelter next to the hospital, because the patients did not meet their burden of proof).

404.   *See* French v. Vix, 21 N.Y.S. 1016, 1023 note (N.Y. Ct. Com. Pl. 1893) (discussing Brennan v. Gellick, a case decided in the superior court of New York City on April 28, 1892).

405.   *See id.*

406.   *See id.*

407.   *See id.*

408.   *Id.*

409.   *See id.*

410.   Harris Stanley Coal & Land Co. v. Chesapeake & Ohio Ry. Co., 154 F.2d. 450 (6th Cir. 1946).

The district court heard testimony from expert witnesses offered by both sides, with like numbers attesting to and denying the alleged danger.[413] The district court ruled in favor of the coal company, reasoning "that the issuance of the injunction would cause great injury to the coal company, [and] confer no comparable benefit on the railroad."[414] Slipping and subsidence caused by the pillar pulling, the court found, was possible and even probable, but there was some doubt as to whether railroad property would in fact be damaged.[415] Regardless, property damage could be remedied with money damages, the district court reasoned.[416] As to the prospect of a landslide occurring at a precise time and place so as to kill people, the district court reasoned that such an outcome "would require a coincidence of events that can hardly be raised to the status of probability."[417]

The Sixth Circuit reversed.[418] The appellate court emphasized that the primary test in equity for an injunction was whether the remedy at law—money damages—would be adequate.[419] Acknowledging low odds of disaster, the court emphasized the magnitude of the possible harm at issue.[420] The court's analysis is highly applicable to the black-hole case:

> If the threatened injury to the railroad right-of-way be envisioned merely as the sliding of some of the surface material of the mountain upon the railroad right-of-way necessitating some expense in its removal and in the repair of the roadbed, we might well say that recovery of damages in a suit at law provides adequate remedy. We have here, however, a railroad over which pass trains bearing passengers and freight. Their daily number is not disclosed by the record, and being but a branch line it may be assumed that the traffic is not heavy. Nevertheless, traffic there is, and the effect of a substantial mountain slide upon a passing train might well be catastrophic. It may be that such disaster could occur only upon a concatenation of circumstances of not too great probability, and that the odds are against it. It is common experience, however, that catastrophies occur at unexpected times and in unforeseen places . . . . A court of equity will not gamble with human

---

411.   *See id.* at 452.
412.   *Id.* at 451.
413.   *See id.* at 452.
414.   *See id.*
415.   *See id.*
416.   *See id.*
417.   *Id.*
418.   *See id.* at 454.
419.   *Id.* at 453.
420.   *Id.*

life, at whatever odds, and for loss of life there is no remedy that in an equitable sense is adequate.[421]

The last three sentences seem particularly fitting to the black-hole case—especially when one considers that the factual context is subsiding earth. Thus, although perhaps quite unusual, courts clearly have the power to enjoin conduct where human life is at issue. What is more, *Harris Stanley Coal* indicates that declining to do so may be reversible error.[422]

## C.  Getting Jurisdiction

Getting jurisdiction over CERN in Switzerland or France, where CERN's campus is located, would be elementary—except for the treaties CERN has established with its host states.[423] The treaties guarantee CERN immunity from legal process,[424] inviolability of its grounds and buildings,[425] and inviolability of its documents and archives.[426]

As is the norm for intergovernmental organizations, CERN is required to make provisions of some kind for settling disputes with private parties.[427] To meet its obligation, CERN is not required to submit to the jurisdiction of any court; instead, CERN is merely required to provide for arbitration or something similar.[428] There are no detailed requirements. The Swiss treaty requires only that CERN "make provision for appropriate methods of settlement of . . . disputes in private law to which the Organization is a party."[429] The French treaty requires CERN to "lay down appropriate rules" for contract disputes and to submit other disputes to arbitration.[430]

Thus far, it appears that no plaintiffs have attempted to avail themselves of an opportunity to undertake arbitration by serving a demand upon CERN. With the laboratory in control of the form of such arbitration, such a demand might well be pointless for objectors.

CERN's immunities do not, however, extend globally. Specifically, no immunities would protect CERN in a U.S. court. Instead, the problem with filing suit in the United States is getting personal jurisdiction.

---

The most fascinating question in this regard is whether a state could exercise personal jurisdiction over a defendant whose only connection to the state was the hypothetical injury from an anticipated tort. For example, could Hawaii exercise jurisdiction over CERN, even if CERN had no Hawaii contacts, solely on the basis that the apprehended harm, should it come to fruition, would obliterate Hawaii? There are strong arguments that jurisdiction should work this way, especially in cases where lack of such jurisdiction would leave the case wholly injudicable.

In reality, however, no such legal machinations are necessary when it comes to the LHC because CERN has left itself open to suit in U.S. courts. In CERN's desire to cut costs for the LHC program, CERN has established sufficient contacts with the United States to provide for personal jurisdiction.

In Illinois, for example, the jurisdiction issue would be clear-cut. Illinois's long-arm statute provides for in personam jurisdiction to the full extent allowed by the U.S. Constitution.[431] The Constitution permits a state to exercise jurisdiction over a defendant even if the defendant does not reside in that state, so long as the defendant has "certain minimum contacts with it such that the maintenance of the suit does not offend 'traditional notions of fair play and substantial justice.'"[432] Sufficient minimum contacts requires that the defendant has "purposely avail[ed] itself of the privilege of conducting activities within the forum State."[433] That availment has occurred in Illinois, because although the collider itself resides in Europe, an important portion of LHC operations are conducted in Illinois[434]—more than enough to satisfy the requirements of minimum contacts and purposeful availment.

One aspect of CERN's Illinois contacts is the LHC Computing Grid ("LCG"). A worldwide network of computers, the LCG is to be used to process and analyze the data gathered by LHC experiments.[435] The LCG is crucial to the LHC project because the LHC's mission of scientific discovery is possible only through the analysis of the enormous amounts of data gathered by the LHC detectors.[436] The primary reason CERN created the LCG was money.[437] When CERN began working on the design for the LHC computing systems in 1999, CERN realized that the necessary computer capacity was far beyond its funding abilities.[438] Some measure of the LHC's computing needs is found in the amount of data it will generate. When active, the LHC will produce about 15 petabytes of data per year.[439] If put on CDs, 15 petabytes require a stack more than 12 miles high.[440]

---

431.    *See* 735 ILL. COMP. STAT. ANN. 5/2-209(c) (West Supp. 2009).

432.    *See* Int'l Shoe Co. v. Washington, 326 U.S. 310, 316 (1945).

433.    *See* Hanson v. Denckla, 357 U.S. 235, 253 (1958).

434.    *See* Erik Sofge, *Large Hadron Collider Turns on Sept. 10, Tests Beam on Weekend*, POPULAR MECHANICS, Aug. 7, 2008, http://www.popularmechanics.com/science/extreme_machines/4276847.html.

435.    *See* Memorandum of Understanding for Collaboration in the Deployment and Exploitation of the Worldwide LHC Computing Grid CERN-C-RRB-2005-01 2 (Apr. 4, 2008), *available at* https://lcg.web.cern.ch/LCG/MoU/Template%20blanks/WLCGMoU_4April2008_blank2.pdf [hereinafter Memorandum of Understanding].

436.    *See* CERN, Aims of the WLCG Project, http://lcg.web.cern.ch/lcg/public/aims.htm (last visited Nov. 30, 2009).

437.    *See* CERN, Distributed Computing, http://lcg.web.cern.ch/lcg/public/distributed.htm (last visited Nov. 30, 2009).

Thus, the LCG exists to utilize computer capacity at sites around the world. Assets and personnel in the United States are slated to handle a quarter of the full nominal data rate, and half of that will be handled by Fermilab in Batavia, Illinois.[441] CERN has also formed a relationship with several American universities for computing services, including the University of Chicago.[442] CERN provides overall management and coordination for the LCG,[443] while receiving around-the-clock grid operations support from Fermilab and the Brookhaven National Laboratory in New York.[444] Stateside staff, though technically employed by their American institutions, are granted the status of "Associate Member of the Personnel of CERN" to the extent they are required to perform work on the CERN site in Europe.[445]

Much less has been found to provide personal jurisdiction. In *Verizon Online Services, Inc. v. Ralsky*, for example, a federal district court held that using a computer network within Virginia was enough for jurisdiction in that state.[446]

Other than the LCG, additional aspects of CERN's operations would provide for jurisdiction as well, including travel by CERN personnel to the United States for various collaborations, such as the fabrication of LHC components in American laboratories.[447]

---

438.   *See id.*

439.   *See* CERN, WLCG Worldwide LHC Computing Grid, http://lcg.web.cern.ch/lcg/public/ (last visited Nov. 30, 2009).

440.   *See* CERN, *supra* note 38, at 45 ("The data flow from all four experiments will be about 700 MB/s, that is around 15 000 000 (=15 PB) per year, corresponding to a stack of CDs about 20 km tall each year.").

441.   *See* Memorandum of Understanding, *supra* note 435, at A8.2 (noting that the target data rate at FNAL is 200 megabytes per second). FNAL is the standard acronym to represent Fermilab in Batavia, Illinois. *See generally* Fermilab, http://www.fnal.gov (last visited Nov. 30, 2009).

442.   *See id.* at A2.4, A2.5.

443.   *See id.* at A3.2.

444.   *See id.* at A6.12. BNL is the standard acronym to represent Brookhaven National Laboratory in Upton, New York. *See generally* Brookhaven National Laboratory, http://www.bnl.gov/world (last visited Nov. 30, 2009).

445.   *See id.* at 9.

446.   *See* Verizon Online Servs., Inc. v. Ralsky, 203 F. Supp. 2d 601, 604 (E.D. Va. 2002).

447.   *See, e.g.*, International Co-operation Agreement Between the European Organization for Nuclear Research (CERN) and the Department of Energy of the United States of America and the National Science Foundation of the United States of America Concerning Scientific and Technical Co-operation on Large Hadron Collider Activities 1997, U.S.-CERN, art. IV, Dec. 8, 1997, 1997 U.S.T. LEXIS 32 [hereinafter U.S.-CERN].

VI. CONUNDRUMS FOR COURTS

Although the LHC provides some unique problems for plaintiffs, the most mind-bending challenges are those that await a judge.

So far, no LHC litigation has reached the merits of the controversy. Perhaps none will. Many courts, perhaps most, would be predisposed to clear such a case off the docket through jurisdictional or procedural means, to the extent doing so is plausible.

Nonetheless, the question of how a court could or would tackle the merits is a fascinating inquiry. The effort creates several puzzles.

Especially intriguing, I think, is to consider how the legal problems posed by black holes mirror the problems they create for physics. Physicists relate that in the vicinity of a gravitational singularity, equations break down, and the known laws of physics seem to fail.[448] Below, with reference to American law, I discuss three lines of legal doctrine that suffer similarly: preliminary-injunction analysis, expert-testimony gatekeeping, and cost-benefit analysis. Developed for a world of automobile accidents, toxic waste, and teratogenic pharmaceuticals, these doctrines all start to break down when confronted with the extreme facts of the black-hole case.

*A.   The Preliminary Injunction Puzzle*

Under American law, a preliminary injunction is a way for a court to order an immediate halt to a specified activity, without the necessity of going through a full course of discovery and trial. The hearing can be held as quickly as 10 days after the defendant has been given notice, and the injunction will last until the completion of a trial on the merits—a process that can take years. Preliminary-injunction requests are common in intellectual-property cases, environmental cases, and with certain other fact patterns, such as impending plans to demolish buildings of historical significance.

To get a preliminary injunction, a plaintiff must make a certain showing under a certain preliminary injunction standard.[449] Different courts articulate the standards differently, and generally speaking, a given court will enunciate alternative formulations, with plaintiffs being the beneficiary of whichever formulation best fits the facts of their case.[450] Across the sundry modes of preliminary injunction analysis, there are two essential lines of inquiry: (1) how likely it is that the plaintiffs will ultimately prevail at trial and (2) how much of a need there is for the requested preliminary relief.[451] There is an inverse

---

relationship between the two factors: the more required under one line of inquiry, the less required under the other.[452]

One of the formulations, which allows for a preliminary injunction even in very speculative cases, requires a showing (1) that "serious questions are raised" and (2) that the hardships that would be caused by the injunction tip "*sharply*" in favor of the plaintiff.[453] Let us examine the LHC case under this standard.

First, compare the hardships. Granting the requested relief would shut down, possibly for years, one of the most expensive, complex, and ambitious scientific undertakings in human history. Further, the discoveries the LHC could enable, which might alter and greatly expand our understanding of the universe, would be removed years into the future. Finally, a preliminary injunction would idle thousands of workers at CERN. That is a lot of hardship.

On the other side of the scales is the Earth and everyone on it being devoured by a black hole.[454]

Result? The black-hole disaster tips the scales well beyond "sharply." No doubt about that.

Now we just need to consider whether "serious questions" are raised. This is where the analysis gets difficult. Human extinction is certainly "serious" in one sense of the word, but this is not, of course, the sense with which we are concerned. What the court must do is determine whether there are questions which ought to be *taken seriously*.

So, should a court take these questions seriously? The plaintiffs say there is a significant chance the world might end. The defendants say the LHC presents "no conceivable danger."[455]

Who is right? Who is wrong? Let us put ourselves in the position of the judge. Our first instinct is to make an independent evaluation of the matter. So, what result do we get when we check the calculations for ourselves?

Of course, that is precisely the problem. We can do no such thing. The subject is utterly recondite. The few people on Earth who are capable of intimately understanding the subject matter form a very exclusive club. Judges and lawyers are not members.

"Modern particle physics is, in a literal sense, incomprehensible,"[456] wrote David Lindley, a former research fellow with the Theoretical Astrophysics

---

452. *See id.* at 11–12 (discussing the "sliding scale").

453. Charlie's Girls, Inc. v. Revlon, Inc., 483 F.2d 953, 954 (2d Cir. 1973) (emphasis added); *see also* Freecycle Network, Inc. v. Oey, 505 F.3d 898, 902 (9th Cir. 2007); Earth Island Inst. v. U.S. Forest Serv., 442 F.3d 1147, 1158 (9th Cir. 2006).

454. One might object that the gross amount of hardship weighed by the court on behalf of the plaintiffs should be discounted by the probability that a black-hole disaster will not come to pass. For a discussion of this kind of mathematical manipulation, see discussion *infra* Part VI.C. The upshot is that the harm of a black-hole disaster is so great, it tends to swamp even tremendous probability discounting.

455. LSAG, *supra* note 179, at cover page.

456. DAVID LINDLEY, THE END OF PHYSICS: THE MYTH OF A UNIFIED THEORY 18 (1993).

Group at Fermilab.[457] "It is grounded," he wrote, "in a sophisticated and indirect mathematical language of fields and interactions and wave-functions."[458] Particle physicists speak, according to Lindley, "in a private gobbledygook understandable only to those similarly initiated."[459]

In particular, the Giddings and Mangano article,[460] the "mammoth"[461] paper at the heart of the scientific controversy, is so complex that CERN's Ellis seemed to think it might not be understandable *"even to reasonably well-educated physicists"* who were not trained in particle physics.[462] What chance, then, does a judge have of thoroughly understanding the matter when considering the matter on a motion for preliminary injunction?[463] Keep in mind that the hearing for a preliminary injunction might be set for 10 days after the commencement of the lawsuit.

Perhaps I might be a kind of yardstick. My ability to pick up and understand the physics in the black-hole case is probably about what one would find on the bench. In the course of writing this article, I have had a great deal of time to educate myself about the science—probably much more time than a judge would have. But while I am now able to follow along with much of the scientific argumentation, that capacity is far short of what would be required for me to comfortably step in and say that one scientific contention is right and another is wrong.

Thus, with the scientific controversy before us, we lawyers are sent looking for our old fallback in such a situation—the place we knew we were headed all along: expert testimony.

---

457. *See id.* on book jacket.

458. *Id.* at 18.

459. *Id.* at 19.

460. *See generally* Giddings & Mangano, *supra* note 211.

461. Ellis, *supra* note 83 (beginning at 2 min.).

462. *Id.* (explaining that the LSAG distilled down the key arguments of the Giddings and Mangano paper to a level similar to the magazine *Physics Today* so that any reasonably well-educated physicist, even if not trained in particle physics, could understand it) (beginning at 3 min.).

463. To be precise, I should note that under some hypothetical facts, one can imagine that a judge might indeed quickly uncover physical questions that could be judged as "serious," and on that basis issue an injunction. But if the scientific arguments of CERN were so transparently unpersuasive, one imagines that public opinion could likely, if not certainly, shut down the LHC without the court's help. Otherwise, one could imagine that a court might judge a question to be serious in the event that CERN simply neglected to respond to some critique that is simple enough to be comprehensible to the layman. Indeed, that has arguably happened with regard to Plaga's paper. *See* discussion *supra* Part IV.H. Barring such scenarios, we are left with the great likelihood that if there were "serious questions" within the equations and data, a judge, even if able to perceive such questions, would not be independently able to decide that they were serious, regardless of how skillfully the plaintiffs' lawyers endeavored to explain them.

## B. *Expert Testimony at the Edge*

LHC defender and theoretical physicist Brian Cox wrote, "Whilst I understand that much of the language of particle physics is opaque, there does come a time when it is worth accepting the views of experts."[464]

That sounds reasonable, but, unfortunately, there is a problem with experts. While one expert will generally give you information you can use, in an adversarial proceeding, you will generally get at least two experts—one on either side of the case. The sum of the testimony then becomes much less helpful. The point was made in a funny sort of way in a note in the *Albany Law Journal* in 1872: "The summoning of expert witnesses by plaintiff and defendant, like the collision of opposing rays of light, ends only in utter darkness."[465]

One way to choose between dueling teams of experts is to identify the experts on one side as having bias. That can be a good tack generally, but it is problematic in the black-hole case: Every expert has a very personal stake in the matter. Generally speaking, the experts are either afraid for their livelihoods or afraid for their lives.

There is, however, another option. Expert testimony can be weeded out before it is even heard. The U.S. Supreme Court has created an entire framework for judges to use when litigants attempt to bring science into the courtroom through expert testimony. Propitiously for the LHC case, the high court's *Daubert* trilogy[466] was formulated specifically to guide trial courts in how to sift out unreliable scientific opinion.

*Daubert* provides that in making a threshold determination of scientific validity, the courts should look to various factors.[467] One consideration is whether the expert's asserted theories are testable, falsifiable, and refutable.[468] A moment's thought reveals that this takes us immediately to a logical absurdity, for it is the testing itself that hangs in the balance of the injunction determination. The theories of Rössler[469] and Plaga[470] can only be confirmed through the obliteration of the court, the parties, and the planet. The theories of the CERN scientists cannot, in the strict sense, be confirmed at all, since incident-free operation of the LHC, even for a period of years, does nothing to confirm its safety on a continuing basis. Of course, CERN's theories are, technically speaking, falsifiable—but only in the unhelpful event that the Earth is destroyed, which renders any hypothetical falsifiability illusory, making it useless as an indication of scientific validity.

---

464.   *See* O'Neill, *Anyone Who Thinks the LHC Will Destroy the World is a Twat*, *supra* note 356. (noting that Brian Cox is "often referred to as the 'rockstar of physics'"); O'Neill, *A Statement By Professor Brian Cox*, *supra* note 356.

465.   *Juries of Experts*, 5 ALB. L.J. 227, 227 (1872), *quoted in* DAVID H. KAYE ET AL., THE NEW WIGMORE: A TREATISE ON EVIDENCE: EXPERT EVIDENCE § 10.2.2, at 336 (2004). Light does not behave this way—the technical explanation is that photons do not obey the Pauli exclusion principle. *See* P.C.W. DAVIES, THE FORCES OF NATURE 79 (2d ed. 1986). The author's sentiment about expert testimony, however, is highly accurate.

466.   Kumho Tire Co., Ltd. v. Carmichael, 526 U.S. 137 (1999); Gen. Elec. Co. v. Joiner, 522 U.S. 136 (1997); Daubert v. Merrell Dow Pharms., Inc., 509 U.S. 579 (1992).

467.   *See Daubert*, 509 U.S. at 593.

468.   *See id.*

469.   *See generally* RÖSSLER, *supra* note 217.

470.   *See generally* Plaga, *supra* note 303.

Yet the problems of testability, falsifiability, and refutability go even deeper—beyond the work of Giddings and Mangano[471] and straight to the heart of contemporary particle physics. Unlike other fields of modern science, such as cell biology or organic chemistry, particle physics has reached a level of theoreticalization such that it is substantially divorced from real-world concepts of testability.

Lindley describes "the inexorable progress of physics from the world we can see and touch into a world made accessible only by huge and expensive experimental equipment, and on into a world illuminated by the intellect alone."[472] Even some within the particle-physics community think that "the trend toward increasing abstraction is turning theoretical physics into recreational mathematics,"[473] an effort that is becoming "ultimately meaningless because the objects of the mathematical manipulations are forever beyond the access of experiment and measurement."[474]

While reproducibility is a hallmark of validity in most scientific disciplines, particle physicists have exempted themselves from its stringent requirements. Experiments are not duplicated in any normal sense because of the expense of the experimental apparatuses.[475] As anthropologist Sharon Traweek explains:

> The meaning of the word *reproducible* here is problematic: no two detectors are alike. No one could get funding to build a copy of another detector and no one would want to try: there would be no credit and influence to be gained. Furthermore, only the group that built the original would have the knowledge to build the copy.[476]

In discussing the indirect evidence relied upon by Carlos Rubbia and CERN physicists in 1981 for claiming discovery of the W and Z bosons, science historians Lloyd Motz and Jefferson Hane Weaver wrote, "One cannot avoid a feeling of uneasiness about this kind of physics, since so much of it is based on the assumed existence of particles that cannot be observed."[477]

Given such a state, it is not clear that any particle-physics testimony should be allowed in the courtroom. The singular and perhaps dubious methodological pedigree of the particle-physics discipline might require its exclusion from the courtroom if it were somehow offered as proof of causation in a personal injury case or to identify the assailant in a murder trial. But the black-hole case is different. Drawing a line in the sand against all particle-physics testimony would be equivalent to judicial abstention. The lack of traditional scientific rigor does not merely cut against the testimony of pro-LHC experts; it cuts against the plaintiffs' experts as well. That is because the only way to make the argument that a particle accelerator could destroy the planet is to do so on the basis of modern physical theory.

---

471.  *See generally* Giddings & Mangano, *supra* note 211.
472.  LINDLEY, *supra* note 456, at 19.
473.  *Id.*
474.  *Id.*
475.  *See* TRAWEEK, *supra* note 247, at 159.
476.  *Id.* (footnote omitted).
477.  LLOYD MOTZ & JEFFERSON HANE WEAVER, THE STORY OF PHYSICS 354–55 (1989).

*Daubert* instructs that another pertinent, though not dispositive consideration is whether the scientific theory at issue has been subjected to peer review and publication.[478] At first blush, this would seem to favor CERN's experts heavily. The Giddings and Mangano article was published by *Physical Review D*, a peer-reviewed journal.[479] Rössler's paper, as he predicted, was rejected by the usual journals.[480] Plaga's paper[481] has not been printed in a journal either. Yet looked at a little bit more carefully, we see that CERN is not really gaining ground here. First of all, Rössler's and Plaga's papers have been published and submitted to peer-review in the sense that they have been made available on the internet, and the internet enables the scientific community to discuss and scrutinize, as well as to memorialize the resulting dialectic. Indeed, that is precisely what happened to Rössler's and Plaga's papers—they were scrutinized and discussed at length.[482] One might argue that this is not the authentic kind of peer review and publication that *Daubert* is talking about. But that is not true. *Daubert* does not extol peer-review and publication because they serve as stamps of approval.[483] Blackmun's opinion lauds peer review and publication because "submission to the scrutiny of the scientific community is a component of 'good science,' in part because it increases the likelihood that substantive flaws in methodology will be detected."[484] It is important to point out that the Court in *Daubert* actually overturned a lower-court decision that excluded testimony on unpublished science that was not subjected to peer review.[485] Moreover, *Daubert* does not refer to "peer-reviewed publication," but rather "peer review and publication."[486] Rössler and Plaga did exactly that. Whether flaws have indeed been revealed is a question that just circles us back to the recondite issue of who is correct on the science.

---

478.   *See Daubert*, 509 U.S. at 593.

479.   *See* Giddings & Mangano, *supra* note 211.

480.   Note, however, that Rössler has published a related paper in a journal concerned with fractals and chaos theory. *See* Otto E. Rossler, *A New Test for E-Infinity Fractal-Spacetime Theory*, 41 CHAOS, SOLITONS & FRACTALS 2291 (2009).

481.   *See generally* Plaga, *supra* note 303.

482.   *See, e.g.*, Giddings & Mangano, *supra* note 211.

483.   *See Daubert*, 509 U.S. at 593.

484.   *Id.*

485.   *See id.* at 584.

486.   *See id.*

A final consideration from *Daubert* is the notion of "general acceptance."[487] *Daubert* teaches that general acceptance among the scientific community, or the lack of such acceptance, may have a bearing on the admissibility inquiry.[488] Here, the "general acceptance" factor heavily favors CERN. Rössler's and Plaga's theories are not generally accepted by the scientific community; CERN's are. So does that end our inquiry? It cannot. The problem is that general acceptance, in this case, causes us to run headlong into a problem of bias that has the effect of rendering general acceptance to be unprobative. All the experts testifying in the LHC cases are going to have demonstrable bias—those in CERN's favor will tend to fear for their livelihoods should they testify against CERN. General acceptance, or the lack thereof, from a community of particle physicists means next-to-nothing when that community itself has a stake in the matter.

Moreover, considering general acceptance in the LHC case, we encounter circularity. If the Rössler and Plaga theories were generally accepted, and CERN's were not, there would be no litigation in which to consider the issues, since LHC construction would have been halted long ago. It is the very fact that the theories are not generally accepted that gives rise to the litigation.

Since it is the conduct of the scientists themselves that is at issue, requiring scientific-expert opinion to be "generally accepted" would be tantamount to making consensus decisions of the scientific community on laboratory-safety issues unsusceptible to judicial review. The very gravamen of the complaint is that the particle-physics community has improperly assessed the risk of that selfsame community's great scientific experiment. General acceptance loses its validatory power when the community is presumptively biased.

It should also be noted that *Daubert* was a repudiation of *Frye v. United States*,[489] in which "general acceptance" was the sine qua non of expert admissibility.[490] Especially when considered together with *Daubert*'s rejection of *Frye*, the logical inappropriateness of the general-acceptance factor makes it clearly unhelpful for sifting through expert testimony in the LHC case.

In the end, doctrine regarding expert-testimony gatekeeping does nothing to help us resolve the dispute.

### C.  *Cost–Benefit Analysis Blows Up*

To illuminate certain legal questions, such as liability for negligence, there is a modern tendency for judges and legal scholars to sometimes undertake the task of using cost–benefit analysis to characterize the facts before them.[491] The

---

487.  *Id.* at 594.

488.  *See id.*

489.  Frye v. United States, 293 F. 1013 (D.C. Cir. 1923); *see also Daubert*, 509 U.S. at 589 (noting "[t]hat the *Frye* test was displaced by the Rules of Evidence").

490.  *See id.* at 1014.

491.  *See* RICHARD A. EPSTEIN, TORTS 129 (1999). For a renowned example of such a case, see *Indiana Harbor Belt Railroad Co. v. American Cyanamid Co.*, 916 F.2d 1174, 1180 (7th exercise is generally traced back to Judge Learned Hand's opinions in two cases involving tugboats: *T.J. Hooper*[492] and *United States v. Carroll Towing Co.*[493] The cases involved questions of, respectively, whether the reasonable tugboat operator would equip each boat with a radio to receive warnings of coming storms[494] and whether the reasonable shipping line would keep a bargee in attendance to mind a moored barge.[495] Hand suggested an algebraic-type analysis for precaution-taking in accident cases,[496] an endeavor that has come to be known as the "Hand formula."[497]

According to the Hand formula, a precaution should be undertaken when the cost of the burden is less than the probability of accident multiplied by the potential loss.[498] Hand's formulation is often expressed symbolically as $B < PL$, where $B$ is the burden, $P$ is the probability, and $L$ is the loss.[499] If, after plugging in the numbers, this inequality is true, then the precaution is reasonable, and the defendant who fails to undertake it will bear the liability under common-law negligence, admiralty law, or the like.

The same sort of analysis translates easily to an injunction context, in which we can ask whether the benefits outweigh the costs. We can calculate the price of risk in a particular endeavor, and then add that to the costs in order to compare the sum to the expected benefit. Using $P$ and $L$ from the Hand formula, the price of risk, $R$, is calculated as:

$$R = PL \qquad \text{(Eq. 1)}$$

Let's try plugging in a few numbers to see what turns up in the black-hole case. First, let us assume that since the Earth and everything on it is the sum of all value for humanity, there is no price worth paying to absorb that loss. Thus, the value of $L$ is infinite.

Next we need to assign a numerical value for the probability. Let us take, for the sake of argument, Giddings and Mangano's assessment that there is "no risk of any significance whatsoever" from the LHC in terms of a planet-devouring black hole.[500] What number do those words suggest? One sensible interpretation is that their numerical equivalent is *zero*. In that case:

---

Cir. 1990) (opinion by Judge Richard A. Posner, fashioning liability rule under Illinois law on the basis of cost–benefit analysis).

492. T.J. Hooper, 60 F.2d 737, 737 (2d Cir. 1932).

493. United States v. Carroll Towing Co., 159 F.2d 169, 170 (2d Cir. 1947).

494. *Hooper*, 60 F.2d at 737.

495. *Carroll Towing*, 159 F.2d at 173.

496. *Id.*

497. WILLIAM M. LANDES & RICHARD A. POSNER, THE ECONOMIC STRUCTURE OF TORT LAW 85 (1987).

498. *Carroll Towing*, 159 F.2d at 173; *see also Hooper*, 60 F.2d at 739.

499. *Carroll Towing*, 159 F.2d at 173; *see also* LANDES & POSNER, *supra* note 497, at 85.

500. *See* Giddings & Mangano, *supra* note 211, at 53.

$$R = 0 \times \infty \qquad \text{(Eq. 2)}$$
$$R = 0$$

This math tells us that the price of risk is zero. This would mean the LHC is a no-lose proposition, and we would not even need to analyze the LHC's benefits to reach the conclusion that no injunction should be issued.

Of course, another sensible interpretation of the phrase "no risk of any significance whatsoever" is that the risk is not quite zero, but is rather a number that is incredibly close to zero. After all, Giddings and Mangano qualified the words "no risk" with the words "of any significance whatsoever."[501] So let's insert an incredibly tiny number in there—something that, in terms of human experience, has "no . . . significance whatsoever."[502] Let's try one in one trillion. In that case:

$$R = (1 \times 10^{-12}) \times \infty \qquad \text{(Eq. 3)}$$
$$R = \infty$$

As you can see, we have obtained a radically different result. By varying $P$ by just the tiniest bit—from zero to very-nearly-almost-but-not-quite-exactly zero, the price of risk has shot up from nil to infinity. Note that we are not using a different assumption for $P$. Quite to the contrary, in both Equation 2 and Equation 3, $P$ equals "no risk of any significance whatsoever." The only difference is in how we translate these words into numbers. In either case, we are taking Giddings and Mangano at their word.

The insertion of infinity into the risk equation causes it to blow up. So, with the goal of getting some usable results, let's reconsider our assumption that the destruction of Earth and the killing of every living thing on it is actually an infinite loss. Judge Richard A. Posner has analyzed the RHIC/strangelet scenario with cost–benefit analysis in the context of whether the U.S. Congress or a hypothetical regulatory regime should shut down the RHIC.[503] In this endeavor, Posner estimated the cost of human extinction, "*very* conservatively" he emphasized, at $600 trillion.[504] He arrived at this number by starting with a value of $50,000 per life and grossing it up.[505] We will come back to scrutinize this estimate in a few moments, but for now, let's fill in the rest of the numbers so that we can compare the costs to the benefits.

Determining the benefits that particle accelerators will provide, Posner wrote, is "difficult, maybe impossible."[506] This is the case, he explained, because particle accelerators are not designed to provide monetary benefit— rather, they are designed to discover the laws of the universe.[507] Nonetheless, he concludes, they have, historically, resulted in some useful technological spinoffs.[508] In consideration of this, Posner "plucked the figure of $250 million a year out of the air," a number which Posner regards as "a gross exaggeration," but one that gives the RHIC the benefit of the doubt.[509]

---

501.   *Id.*
502.   *See id.*
503.   *See* POSNER, *supra* note 109, at 187–96.
504.   *Id.* at 141.
505.   *See id.* at 168.
506.   *Id.* at 142.

We can determine the net benefits of the RHIC according to the following, where $B_{R\&D}$ is the earned research-and-development benefit ($2.1 billion over 10 years according to Posner, which is discounted to present value),[510] $C_{C\&O}$ is the costs of construction and operation over the RHIC's 10-year lifespan (Posner's discounted number is $1.7 billion),[511] $P$ is the probability of catastrophe (Posner uses 1 in 10 million per year, multiplied by 10 years),[512] $L$ is the price of extinction ($500 trillion after discounting), and $B_{net}$ is the net benefit. The equation is:

$$B_{net} = B_{R\&D} - C_{C\&O} - PL \qquad\qquad \text{(Eq. 4)}^{[513]}$$

With Posner's numbers plugged in, we can calculate the result:

$$B_{net} = \$2.1 \text{ billion} - \$1.7 \text{ billion} - (((1 \text{ in } 10 \text{ million}) \times 10 \text{ years}) \times \$500 \text{ trillion}) \qquad \text{(Eq. 5)}$$

The net benefits calculate out to a negative net benefit of $100 million.[514] On this basis, Posner suggests the RHIC experiment may not be worth running.[515]

Posner's analysis is problematic for several reasons. First, it lacks any robustness. The fragile nature of the analysis is easy to demonstrate. Note that Posner's value for yearly research and development gains was plucked "out of the air," and his value for human lives is derived from layers of assumption.[516] Thus, it seems entirely fair to pick out somewhat different, though similar, numbers and solve the equation again. The result should be just as meaningful.

For R&D, let us suppose that the gains from the RHIC will be $350 million per year—another number plucked out of the air, but with no lesser claim to validity. After all, one good patented invention can be worth billions. That fixes $B_{R\&D}$ at about $3.0 billion, after being multiplied by 10 years and discounted.

---

For human lives, let us pick $35,000 per life—a number that is entirely within the same realm of supposition. Grossed up in the same way Posner did with $50,000 and discounted, we get around $180 trillion.[517] Using these numbers:

$B_{net}$ = $3.0 billion – $1.7 billion – (((1 in 10 million) × 10 years) × $180 trillion)                                                              (Eq. 6)

The net benefits calculate out to $1.1 billion in positive gains for the RHIC program. That makes it look like a smashing success. Not a boondoggle in the slightest. Using equally reasonable numbers, the result came out wildly different than Posner's. How did that happen?

In Posner's defense, what he was trying to do was show that even using estimates that gave RHIC the benefit of the doubt and then some, the RHIC revealed itself to be a waste of money. Regardless, however, our re-running of the calculations has shown that the number-crunching lacks meaningfulness. If attention is paid to the significance of the numbers chosen, taking into account a reasonable margin of error for each, it becomes clear that the values are entirely inadequate for the task of deriving any kind of significant result.

Now, let us come back to Posner's estimate of the value of cost of human extinction at $600 trillion.[518] Is this absurd? Posner does not think so. According to him, "Valuing human lives is not . . . quite so arbitrary a procedure as it may seem. It sounds like an ethical or even a metaphysical undertaking, but what actually is involved is determining the value that people place on avoiding small risks of death."[519]

Working with people's willingness to pay to avoid low-probability risks of death, Posner arrives at a price per human life of $50,000.[520] That number in hand, he then multiplies it by the world population of six billion.[521] This would give us a human-extinction cost of $300 trillion, but since Posner has not otherwise accounted for future lives, he provides for this omission with a "crude adjustment" of doubling the product. The result is $600 trillion.[522]

Now, wait a minute. Why did Posner calculate the worth of a human life based on a person's willingness to pay for *low-probability* risks of death? Why not *high-probability* risks? Posner notes that economists have looked at data on wages for various occupations with differential risks of death, and, based on that data, calculated the premiums demanded by workers for hazardous jobs.[523] This exercise shows that workers tend to value their own lives at something around $5 million.[524] So why did Posner take it down to $50,000? He notes that

517.    Posner's calculation in this regard is discussed a few paragraphs below.
518.    POSNER, *supra* note 109, at 141.
519.    *Id.* at 165.
520.    *Id.* at 166–68.
521.    *Id.* at 169.
522.    *Id.* at 169–70.
523.    *Id.* at 165.
524.    *Id.* at 166.

willingness to pay to avoid risks does not appear to be related in a linear way to the probabilities of risk.[525] For instance, he notes, if the risk of death were not one in a thousand but one in *two,* people might well demand an infinite amount of money to be exposed to such a risk.[526]

By the same token, Posner figures, at the other end of the asymptotic curve, willingness to pay will fall faster than probabilities of death, to the point where, he conjectures, people might demand zero compensation for bearing truly miniscule risks.[527] Thus, Posner picks $50,000 as what he considers to be a likely value based on the order of magnitude of risk involved in a planet-ending particle-accelerator catastrophe.[528]

But there is a serious problem with Posner's argument. It makes no sense to scale up from $50,000 for one person to $600 trillion for everyone. Why not? It is as simple as this: A human life does not vary in value with the probability that it will be extinguished. Indeed, how can it? Dead is dead. The quality of death is precisely the same whether a person dies from a low-probability accident or a high-probability accident.

This point is easy to understand if we pause to reflect that probability itself is only a mathematical way of representing ignorance.[529] Suppose you pick up a peach and mull over whether or not to eat it. Let us say there are stated odds of one in one billion that the peach is contaminated with a lethal strain of E. coli. Now, those may be the odds, but in reality, the peach either has the E. coli, or it doesn't. Once it is tested or eaten, the probability will become either 100% that it is contaminated or 100% that it is not. Whether the *ex ante* odds of E. coli contamination were stated at one in a billion or one in three, once you've died from eating a contaminated peach, you are 100% dead.

The case of the LHC and black holes is the same. The physical laws of the universe are either configured in such a way that proton-proton collisions at 14 TeV can create dangerous black holes, or the universe is not so configured. The probability—to the extent anyone would venture to posit it—would only be a representation of our relative ignorance on the subject.

The fact that an individual's compensation demands might decrease in a non-linear way with decreasing risk—a supposition I am not taking issue with—does not indicate that resulting deaths are any more or any less costly; it only shows that people are irrational in valuing risk. Irrationality in dealing with risk is nothing new. Posner, himself, has acknowledged it.[530] And it has been the subject of lengthy treatment by Cass Sunstein.[531] A great advantage of cost–benefit analysis is that it helps us to analyze choices in some degree of shelter from irrationality.[532] We lose that advantage by using differential life-valuation for differential probabilities of harm. The adjustment only incorporates the irrationality into the equations.

---

525.   *Id.*

526.   *Id.*

527.   *Id.*

528.   *See id.* at 166–68.

529.   Since the instant subject matter concerns particle physics, I should mention that while probability is, in the general case, a mathematical way of representing ignorance, that is not the case in quantum mechanics. According to quantum theory, the probability of events such as radioactive decay or atomic photon emission is one that is described by probabilities that are fundamental, not based on human ignorance of underlying conditions. *See* FORD, *supra* note 154, at 114.

Undoing that part of Posner's analysis, let us go back and take a standard, non-adjusted value for human life based on economists' work: $5 million.[533] Now let's try the calculation again, multiplying the value of one human life, $V_{\text{life}}$, by the human population of Earth, $H_{\oplus}$.

$$L = V_{\text{life}} \times H_{\oplus} \qquad\qquad\qquad (\text{Eq. 7})$$
$$L = (\$5 \times 10^{6})(6 \times 10^{9} \text{ people})$$
$$L = \$30 \times 10^{15}$$

The value of $5 million multiplied by a population of six billion people is $30 quadrillion. This cost–benefit analysis looks extremely unfavorable for particle-accelerator projects.

But wait—we have not valued future lives. Let us use a recently calculated value for the worth of one additional year of human life for one person: $129,000.[534] We have roughly five billion years to go until the Sun expands into Earth's orbit and cooks the planet.[535] Assuming the population stayed at around six billion people,[536] we can calculate human-extinction damages, or loss, $L$, where $V_{\text{yr}}$ is the value of one human life per year, $H_{\oplus}$ is the human population of Earth, and $Y_{\oplus}$ is the number of years the Earth has left, as follows:

$$L = V_{\text{yr}} \times H_{\oplus} \times Y_{\oplus} \qquad\qquad\qquad (\text{Eq. 8})$$
$$L = (\$1.29 \times 10^{5})(6 \times 10^{9} \text{ people})(5 \times 10^{9} \text{ yr})$$
$$L = \$3.87 \times 10^{24}$$

---

530.   *See* POSNER, *supra* note 109, at 11.

531.   *See, e.g.*, CASS R. SUNSTEIN, RISK AND REASON: SAFETY, LAW, AND THE ENVIRONMENT 28–52 (2002).

532.   *See, e.g., id.* at 39, 49, 292.

533.   POSNER, *supra* note 109, at 166; *see* SUNSTEIN, *supra* note 531, at 222, 224.

534.   *See* Kathleen Kingsbury, *The Value of a Human Life: $129,000,* TIME, May 20, 2008, http://www.time.com/time/health/article/0,8599,1808049,00.html (last visited Nov. 30, 2009); Chris P. Lee et al., *An Empiric Estimate of the Value of Life: Updating the Renal Dialysis Cost-Effectiveness Standard*, 12 VALUE IN HEALTH 80, 86 (2009) (valuing one year of human life at $129,090).

535.   *See* Appell, *supra* note 257, at 24.

536.   Various scholars have reckoned that Earth's natural carrying capacity for humans is higher than six billion, and various scholars have reckoned it lower. Thus six billion is a reasonable number to use here.

The result is $3.87 septillion. A very large amount of money indeed.

But hold on, there is a big gap in our analysis. What if many of those people are and will be completely miserable? Just because we spend money to avoid death does not mean, in the grand scheme of things, that life provides a net benefit. If the Earth suddenly collapsed into an infinitely dense point, there would be no more pain, no more suffering, no more hunger, and no more grieving. Depression and anguish are, of course, reasons people commit suicide. It is often the case that if someone commits suicide, there are people left behind who grieve. Thus, there is a social cost to death.[537] But in the future, if what is left of Earth lies at the center of a marble-sized black hole, there will be no soul left to shed a tear.

This conception of human death as essentially lossless may sound heartless, but it is an idea that is enshrined in the American common law of torts. The estate of a dead person who is killed by negligence is entitled to no damages—at least not on account of the person being dead. There are wrongful-death statutes that may allow suffering family members to recover.[538] There are also survival actions whereby the estate may recover for pain and suffering experienced by the decedent in the moments before death.[539] But there are no damages for death itself. In fact, it is generally the case that if someone kills a person who has no family or loved ones, provided the person's death is instantaneous, the wrongdoer will not end up liable for even a single penny. Absent special circumstances, death is simply not a redressable injury under American tort law.

Thus, maybe the downside of a particle-accelerator disaster that destroys the planet—assuming it is quick—is nothing.

Although Posner said that valuing human lives was not "an ethical or . . . metaphysical undertaking,"[540] in fact, it is—at least when it comes to the end of the world.[541]

The bottom line? Cost–benefit analysis sheds little light on the situation. To illuminate our way out of the black-hole case, we will need something new.

## VII. ESCAPING THE VORTEX

For courts, the challenges in providing fair and meaningful adjudication in a particle-physics doomsday-injunction case are legion. But if the judiciary surrenders to these difficulties and refuses to involve itself in such disputes—for instance, by giving short shrift to plaintiffs' claims or using discretionary aspects of civil procedure as a pretext for bowing out—then the judiciary renders a class of consensus judgments within scientific communities effectively injudicable. In that event, the rule of law is lost. Such a result seems especially unacceptable when the alleged harm is the destruction of the Earth.

---

537. Posner acknowledges this. *See* POSNER, *supra* note 109, at 166 .

538. *See, e.g.*, N.D. CENT. CODE § 32-21-01 (1996).

539. *See, e.g.*, Nelson v. Dolan, 434 N.W.2d 25, 26, 30 (Neb. 1989).

540. *See* POSNER, *supra* note 109, at 165.

541. There are good arguments—which I do not discuss here—for assigning monetary values to lives for the purposes of regulatory trade-offs and budgeting government spending. But such arguments do not apply to balancing prospects for the extinction of humanity.

Yet if the judiciary plows ahead and issues an injunction in such cases—deciding that we are better safe than sorry when the merits of a case are obscured by mathematical or scientific complexity—then courts would be transformed into marionettes—manipulable by frivolous objectors into halting any scientific undertaking that is complicated enough to be opaque to the layperson. That seems unacceptable as well.

Resolving this quandary is important. As humankind continues to advance in scientific and technological achievement, there will be more and more opportunities to consider what exotic dangers might be lurking around the corner. If and when the titans of science and industry find themselves at odds with bystanders about what constitutes acceptable risk to the environment and the human species, lawyers and judges are the citizens' bulwark. When it is up to judges and lawyers to save the world, will they be ready?

In this portion of the article, I suggest analytical methods that will allow courts to deal meaningfully and responsibly with the surreal challenges of having Armageddon on the docket.

But let us pause for a moment for a reality check: Is it really plausible that a group of extremely smart, highly trained, non-sociopathic scientists and engineers could overlook fatal flaws in a wildly expensive, high-stakes project and thus cause a deadly catastrophe? Of course it is. It is not only plausible, but it has already happened multiple times. Examples include the space shuttle *Columbia* disaster, the space shuttle *Challenger* disaster,[542] and the disastrous Castle Bravo thermonuclear bomb test.[543] Non-lethal examples include many unmanned space-program mishaps, such as the Mars Climate Observer and Europe's heavy-lift Ariane 5 rocket.[544] In the realm of biology and medicine, there are numerous examples of drugs that were once deemed safe, only to be later removed from the market.[545]

What these and other examples teach us is that there are ways that courts can evaluate seemingly recondite technologies and scientific controversies. The trick is to focus on the comprehensible factors—namely, the human factors. This is possible because every scientific failure takes place within a human context. A powerful example comes from the *Columbia* disaster. In its final report, the Columbia Accident Investigation Board concluded that foam debris from the external tank, which fatally punctured the orbiter's heat shield, was only part of the cause: "In our view, the NASA organizational culture had as much to do with this accident as the foam."[546]

---

542. *See* 1 COLUMBIA ACCIDENT INVESTIGATION BOARD, COLUMBIA ACCIDENT INVESTIGATION BOARD REPORT 200 (2003) [hereinafter CAIB].

543. *See* Toby Ord, Rafaela Hillerbrand, & Anders Sandberg, *Probing the Improbable: Methodological Challenges for Risks with Low Probabilities and High Stakes* 9, http://www.fhi.ox.ac.uk/__data/assets/pdf_file/0006/4020/probing-the-improbable.pdf.

544. *See id.* at 7.

545. *See, e.g.*, Joel Lexchin, *Drug withdrawals from the Canadian market for safety reasons, 1963–2004*, 172 CMAJ 765 (2005) (available at: http://www.cmaj.ca/cgi/reprint/172/6/765.pdf) and online appendix (available at: http://www.cmaj.ca/cgi/data/172/6/765/DC1/1).

While courts may not be well equipped to settle scientific disputes on the scientific merits, courts are quite well equipped to look at the human aspects in a prospective catastrophe and render a decision on an injunction request on the basis of those factors.

Human failings in science come in many sorts. Many species of error may be quite innocent, such as those caused by faulty data, miscalculations, mistaken scientific theory, or conceptual blunders. In addition, well-meaning people may be subject to psychological effects and aspects of organizational culture that cause flawed decision-making in scientifically complex contexts. A variety of behavioral schema may help to explain such shortcomings, including groupthink, cognitive dissonance, confirmation bias, and unavailability bias.

At other times, the sources of scientific error are less innocent and may be rooted in willful blindness or conscious self-interest. There are several historical examples. An icon of corrupt science is the cigarette industry's history of bold assertions that its products were not addictive and did not pose health hazards. Global climate change is another relevant example. Today, global warming is no longer the object of serious controversy, but political pressure and industry self-interest arguably had a hand in delaying consensus about the existence of a real threat, thus putting the project of abating greenhouse gas emissions years behind where it would have been otherwise.

Let me be clear that my point in providing these examples is not to suggest anything about the probability of a black-hole disaster. For instance, I am not contending that a black-hole disaster is as likely as a space shuttle mishap. Rather, my point is that society's experience with disastrous scientific and engineering failures shows that deciding an LHC-injunction request, or a similar hypothetical case, may be done fairly if we avail ourselves of the right analytical tools.

Thus, in evaluating whether there are the sort of "serious questions" that justify a preliminary injunction, and whether there is the requisite sort of risk so as to justify a *Harris-Stanley*-type finding of negligence,[547] courts should conduct analysis on a higher level. Using a kind of meta-analysis, courts should gauge the risk that scientific judgments are wrong. Relevant subjects of inquiry include organizational culture, group politics, and psychological context. The particular aspects of scientific arguments should also be scrutinized on a meta level. Relevant issues here include the newness of underlying theory, the complexity of the chain of argument, the likely reliability of underlying data, and so on. Also relevant is what history has to say about the durability of pronouncements made in the field.

---

546.   CAIB, *supra* note 542, at 97.
547.   *See* discussion *supra* Part V.B.

Below, I discuss four categories of meta-analysis that could be applied to the LHC case. These categories are not strictly distinct—there is considerable overlap among them—but the breakdown should be helpful for approaching the problem juristically. The first two categories focus on the scientific work. The second two categories focus on the scientists.

*Defective theoretical groundings*—This category concerns the potential that the scientific theory upon which the safety assurances are based may be defective. This is macro-scale scientific error—the kind of failing that inures to the scientific enterprise as a whole, in its current state of knowledge, with regard to its capacity to provide a reliable basis for accurate risk assessments.

*Faulty scientific work*—This category regards micro-scale scientific error: miscalculations, inaccurate observations, flawed assumptions, conceptual mistakes, and the like. This is error that is attributable to the work of individual scientists or research groups.

*Credulity and neglect*—This category encompasses innocent or at least unintentional mistakes in decision-making and risk assessment because of cultural, psychological, or sociological factors. Relevant concepts here include groupthink and various cognitive biases.

*Bias and influence*—This category concerns the potential for non-innocent errors motivated by self-interest and ambition.

These categories are meant only to carve up the analysis into workable portions. In a case presenting the question of prospective injury from a complex scientific or technological undertaking, a judge could decide on the basis of any one of these categories, or some combination of more than one, that an injunction is warranted.

Below, I describe these categories in some detail. Also, I do some preliminary work of applying them to the facts of the LHC case.

As will become clear, my preliminary assessment looks unpromising for CERN. While it seems absurd, in the abstract, that a group of apparently normal people could risk the entire planet in the course of carrying out a science experiment, the prospect does seem distinctly plausible once one takes a look at the details. Such a disaster is not likely, to be sure, but it does appear plausible enough to give one pause. When it comes to the ultimate question—whether the LHC be stopped—I am agnostic. That decision, to my mind, should be for a judge—one with the benefit of being able to take testimony, compel discovery, review evidence, and hear carefully presented argument from both sides. Society is well served by courts that can undertake these means to settle controversies. Using the tools of civil procedure, courts can command a far superior vantage point on the controversy than I can by reviewing published papers and publicly available documents. Nonetheless, my preliminary review does, in my view, suggest that the controversy deserves sober consideration.

### A.  *Analyzing the Potential for Defective Theoretical Groundings*

Sometimes, despite their best efforts, scientists just turn out to be wrong. In making this observation, we should distinguish between two sorts of scientific failings. One, which is a necessary part of progressing science, is a knowing failure to comprehend something. This is a situation in which science does not know the answer to a certain question, and *science knows that science does not know the answer.* As an enterprise starting from ignorance and seeking knowledge, science must, necessarily, encounter questions to which, at least at some early point, it does not have the answer. Such a situation is non-problematic.

A second kind of failing is the one that we are interested in here, which I am calling "defective." In this situation, *science thinks it knows the answer, but the answer is wrong.* Such error is not a necessary element of the scientific enterprise, but it happens nonetheless.

Applying this line of analysis to the black-hole case, we should consider whether state-of-the-art theoretical physics is inadequate to the task of making a trustworthy prediction that the LHC is safe. In other words, we need to try to gauge the possibility that the science underlying the exclusions of disaster at the LHC may ultimately be wrong.

Outside of the context of accelerator-safety issues, physicists can be quite candid about science's ability to get things wrong. In describing what theoretical physics has to say about some of the discipline's most fundamental questions, Stephen Hawking wrote, "Someday these answers may seem as obvious to us as the earth orbiting the sun—or perhaps as ridiculous as a tower of turtles. Only time (whatever that may be) will tell."[548]

It is important to note that in saying this, Hawking was not speaking about the science underlying the conclusion that the LHC is safe. Quite to the contrary, Hawking, a staunch supporter of the LHC, has stated a firm belief that the LHC is "absolutely safe." In fact, he has gone even further, saying the LHC is vital to the survival of human race.[549] Speaking plainly, he also predicted that the production of black holes at the LHC would net him a Nobel Prize.[550] But Hawking's statement about the fallibility of science reflects a broader truth— one that is inescapable: Scientific theory that seems unassailable in one era may seem naïve in the next.

---

548.   HAWKING, *supra* note 76, at 5.

549.   *Id.* ("Prof[essor] Hawking said the £4.4bn machine, in which scientists are about to recreate conditions just after the Big Bang, is 'vital if the human race is not to stultify and eventually die out.'"). *But cf.* HAWKING, *supra* note 76, at 18 (saying that a fully realized unified theory of nature "may not aid the survival" of the human species or "even affect our lifestyle").

550.   Jon Swaine, *Stephen Hawking: Large Hadron Collider Vital for Humanity*, THE TELEGRAPH, Sept. 9, 2008, http://www.telegraph.co.uk/news/2710348/Stephen-Hawking-Large-Hadron-Collider-vital-for-humanity.html (quoting Hawking as saying, "If the LHC were to produce little black holes, I don't think there is any doubt I would get a Nobel Prize, if they showed the properties I predict").

Arguments made from this premise of skepticism could, of course, quickly get out of hand. It would not be defensible, for instance, to argue that because scientists once thought the Sun revolved around the Earth, today's scientists might be just as mistaken about the LHC. Such an argument proves too much. Not only would it permit a total halt to any challenged experiment, it would reduce all of science to a pile of distrusted mush. In such a case, there would be no point in taking any arguments seriously, whether they are for or against LHC operation.

If an analysis of the fallibility of scientific theory is to be of use to a court, then what we need are ways to bring some principled, discriminating skepticism to bear.

A good starting place is the temporal dimension of scientific theory. The longer a theory persists, the more confidence it deserves. Conversely, if theory is new, it should be afforded less confidence. A matter related to the longevity of theory is the pace of relevant theoretical work. If a theory is the direct subject of back-and-forth papers arguing its merits, the theory deserves, for the time being, less confidence. That is, in so far as the theory is the topic of an active debate, outside observers of that debate, such as courts, should remain skeptical. Thus, a court evaluating the LHC case, or a similar case, should consider the newness of theory upon which the safety analysis is based and the pace of theoretical papers being published about such foundational theory.

In the case of the LHC, there are facts suggesting that theory is evolving too rapidly to form a firm foundation for making risk assessments. CERN's John Ellis has even said that physics relating to microscopic black holes is "a fast moving subject."[551]

Indeed, Don Lincoln, particle physicist and steadfast LHC defender, is candid about science's limitations. "We simply don't know how gravity works in the realm of the ultrasmall," he wrote.[552] In forecasting what might result from the LHC's experiments, Lincoln wrote, "Of course, there can always be surprises. This is the research frontier, after all."[553] Yet, when the specific question of LHC safety arises, Lincoln has resolute confidence, saying the LHC poses "precisely zero risk" and that a planetary disaster is "*impossible.*"[554]

Courts may, with some probing, determine that there is some unsettling inconsistency in the persistent duality whereby particle physicists espouse general skepticism of their craft yet maintain perfect, or nearly perfect, confidence on safety issues. Whether this apparent contradiction should be troubling appears to be answered by a consideration of scientific precedent. The history of particle-collider safety assurances contains a quick succession of flip-flops on theory that necessitated rethinking prior conclusions. This history suggests that there are good reasons to find the current round of safety guarantees less than completely settling. A chart of the history of scientific conclusions and safety assurances is provided in Table 1.

---

551. Ellis, *supra* note 83 (beginning at 10 min.).

552. LINCOLN, *supra* note 48, at 2.

553. *Id*. at 62.

554. *Id*. at 2 (emphasis in original).

| **SAFETY ARGUMENT AGAINST BLACK-HOLE SCENARIO** | | | **UNDERMINED BY EVOLVING THEORY** | | **ULTIMATE DISPOSITION OF SAFETY ARGUMENT** |
|---|---|---|---|---|---|
| 1999 | Accelerators for the foreseeable future will not be powerful enough to create black holes.[555] | → | 2001 | Theorists demonstrate that with extra dimensions, the LHC has the energy required to synthesize black holes.[556] | → *Abandoned:* CERN acknowledged the need for a new examination of potential hazards, since, under new theory, black holes "will be produced."[557] |
| 2003 | Hawking radiation will cause accelerator-produced black holes to evaporate.[558] | → | 2004 | Unruh, a respected pioneer of black-hole evaporation theory, calls the theory into question.[559] | → *Abandoned:* CERN declined to continue resting its safety argument on black-hole evaporation.[560] |
| 2008 | Under some scenarios, synthetic black holes will grow too slowly to be a threat. Under the remaining scenarios, dangerous black holes are excluded on the basis of empirical observation of specified white dwarf stars.[561] | → | | *???* | → *???* |

**Table 1:** Chart of arguments advanced against black-hole disaster scenarios to demonstrate particle-accelerator safety.

The chronology belies the characterization by CERN's Scientific Policy Committee that the basis for safety claims is "firmly established theory."[562] To the contrary, the sequence of quickly shifting theory and opinion within the scientific community would seem to indicate that the opinions reflected in the LSAG report,[563] along with the Giddings and Mangano paper[564] upon which it is based, are too new to be relied upon for a real-world conclusion that there is no risk of catastrophe. In addition, the Giddings and Mangano paper appears to be complex even by particle-physics standards. Even if accepted at first by the scientific community, there would seem to be no guarantee that its layers of argument will stand the test of time—despite the unequivocal nature in which such arguments have been posited.

---

555.	*See supra* Part IV.B.
556.	*See supra* Part IV.C.
557.	*See* LSSG, *supra* note 180, at 11–12; *see* Part IV.D.
558.	*See supra* Part IV.D.
559.	*See generally* Unruh & Schützhold, *supra* note 206, at 2, 11; *see supra* Part IV.D.
560.	*See supra* Part IV.G.
561.	*See supra* Part IV.F.
562.	*See* CERN Scientific Policy Committee, *supra* note 232, at 4.

Steven Giddings in particular has been cited for persistent mistaken understanding. Leonard Susskind, a friend of Giddings, has recounted Giddings's "stubborn attachment to wrongheaded ideas" in the 1990s when Giddings and co-authors made a mistake of theory regarding Hawking radiation, despite having conducted an "elegant mathematical analysis."[565]

"Certitude is not the test of certainty," wrote Oliver Wendell Holmes, Jr. in his essay *Natural Law*.[566] "We have been cock-sure of many things that were not so."[567]

## B.  *Analyzing the Potential for Faulty Scientific Work*

Even where the big picture of scientific theory is secure and justly relied upon, a broad spectrum of smaller-scale errors can arise in the course of scientific work, and these errors can drive researchers to make invalid conclusions. Examples of such micro-scale errors include calculation mistakes, observational errors, and unwarranted assumptions. Such fatal errors might be committed directly by the scientists writing the report under scrutiny, but such errors may also come from other sources. For instance, if scientists rely on data or conclusions provided by other scientists, then material mistakes in those underlying reports will invalidate any conclusions built upon them. Similarly, software bugs can undermine scientific claims—for example, conclusions drawn from computer models are undermined if the software used to construct those models contains defects.

Thus, the potential for these kinds of scientific error forms a paradigm for meta-analysis of scientific claims that is distinct from the potential for defective theoretical groundings, as discussed above.

So, when it comes to the black-hole case, we ought to ask how likely is it that the data relied upon for the safety assessment is inaccurate. This is not to suggest that experts should be hired to comb through underlying data to look for flaws. It is a given that in a case like that of the LHC, because the science is so complex, no expert can be counted upon to uncover all possible mistakes— after all, whatever mistakes were made, were made by experts themselves.

---

The point of considering the potential for small-scale error in scientific work is understanding that scientists' conclusions about risk cannot be taken at face value. Their assurances must be discounted by the probability of their own error. For example, Giddings and Mangano conclude that there is no conceivable risk.[568] But it does not follow that LHC risk is zero: An accurate assessment of risk must include the possibility that Giddings and Mangano themselves are mistaken.

Such a meta-analysis has been proposed by a team of philosophers from Oxford University—Toby Ord, Rafaela Hillerbrand, and Anders Sandberg: "When an expert provides a calculation of the probability of an outcome, they are really providing the probability of the outcome occurring, given that their argument is watertight."[569]

This line of inquiry might strike scientists as unfair. "If I've made a mistake," a scientist might argue, "then point it out to me."

That retort is certainly fair in the context of a debate between scientists. Hypothesizing unspecified phantom flaws to introduce doubt into someone else's work would surely be dubious at a roundtable of physicists. But the issue of LHC risk is not a purely academic debate—and the same rules do not apply. In the context of a policy debate or legal dispute where scientific arguments are used to justify a real-world course of action that is allegedly dangerous, it is improper to take all scientific arguments on their own terms without subjecting them to a higher level of scrutiny.

Moreover, in the case of the Giddings and Mangano piece,[570] meta-level scrutiny seems especially warranted because the complex, technical nature of the paper makes it largely opaque to all but a small community of highly trained scientists.[571] And the need for scrutiny is reinforced when one considers that those qualified scientists are, as a general matter, invested in careers where there may be considerable pressure to fall in line with CERN's safety assessments.

In addition, there is a compelling mathematical reason that meta-level scrutiny of the scientific work is needed in the case of the LHC. Ord, Hillerbrand, and Sandberg demonstrated that when it comes to ultra-low probability risks where the harm alleged is catastrophic, the probability reported by the expert ends up being trivial when the probability of the expert's error is taken into account.[572] Why? The overall probability of disaster occurring is a product of the odds provided by the expert *and* the chance that the expert is wrong.[573] Where the probability given by an expert is extremely low, for

---

568.  *See generally* Giddings & Mangano, *supra* note 211.

569.  *See* Ord et al., *supra* note 543, at 1.

570.  *See generally* Giddings & Mangano, *supra* note 211.

571.  *See* Ellis, *supra* note 83 (beginning at 3 min.).

572.  *See* Ord et al., *supra* note 543, at 2–5.

573.  *See id.* Note that I am simplifying. To be precise, the probability of disaster is the probability provided by the expert multiplied by the probability that the expert is right, plus the probability that the expert is wrong multiplied by the probability of disaster given that the expert instance one in a billion, the chance of human error dwarfs the expert's estimate of risk.

Problems constituting faulty scientific work can be sorted into various categories: flawed arguments, flawed calculations, observational error, and errors in models, assumptions, and conceptual thinking. A consideration of each category illustrates the ways in which well-meaning scientists can produce defective conclusions.

*Flawed arguments*—Ord, Hillerbrand, and Sandberg note that flawed arguments are not rare.[574] They cite studies of the MEDLINE life-sciences database showing that more than 6.3 in 100,000 papers are retracted.[575] This number may seem small, but when the downside of a mistake is the destruction of Earth, such a number is clearly quite significant.

Further, the Ord team argues that this number is conservative for a number of reasons, including that researchers can be expected to resist issuing a retraction unless compelled to do so.[576] Therefore, the number might provide a good floor for estimating the probability that Giddings and Mangano's assessment of zero risk is wrong. On the other hand, CERN might argue that the MEDLINE retraction rate is a poor comparator because Giddings and Mangano's work has received more scrutiny, because it is simpler than the average MEDLINE paper, or for any number of other reasons. Such arguments might be quite valid, and a court should carefully consider them.

*Flawed calculations*—Separate from the issue of flawed arguments is a question of flawed calculations. Ord's team uses statistics and examples to show the importance of this category of error. Calculation mistakes in hospital drug charts cause dosage errors at a rate of about one to two percent—despite an obvious need to be careful when lives are on the line.[577] Academicians, though they have more time to root out error, also make calculation errors. A study of *Nature* and the *British Medical Journal* uncovered flawed statistical results occurring at a rate of about 11%.[578] Software problems are behind mathematical mishaps in many cases.[579] A cascade failure caused by errant software code led to the 1996 explosion of an Ariane 5 rocket launched by the European Space Agency.[580] NASA lost its Mars Climate Observer in 1999

---

is wrong. *Id.* Put in mathematical terms: where $X$ represents the occurrence of catastrophe, and $A$ represents the expert's argument being sound, then the probability of catastrophe, $P(X)$, is given by the formula:

$$P(X) = P(X|A)\, P(A) + P(\neg A)\, P(X|\neg A)$$

The symbol "|" is read as "given" and the symbol "¬" is read as "not." *See id.* at 3.

574.    *See id.* at 4.
575.    *Id.*
576.    *See id.*
577.    *Id.* at 7.
578.    *See id.* at 7.
579.    *Id.*
580.    *Id.*

because a piece of software used English units of measurement while other software used metric.[581] Another software problem led to retraction of five scientific papers on protein structure.[582] And, strangely enough, physicist Don Lincoln, in his laudatory book about the LHC, was off by several orders of magnitude in translating light years to miles and kilometers.[583]

***Observational error***—Another issue, with special relevance for the LHC case, is the possibility of observational error. The ultimate conclusion that the black-hole scenario is excludable rests upon the existence of eight white dwarf stars with certain attributes of age, mass, and magnetic field strength.[584] Clearly, there is some chance that those astronomical observations are in error.

For five of the eight white dwarfs, the only cited observational reference is a single paper published in 2007.[585] Thus, if a flaw were found in that paper or that reference were somehow impeached, the LHC safety argument funnels down to reliance on only three observed stars.

One reason to be somewhat skeptical of the star observations is a bedrock concept from the common law of torts: foreseeability. No astronomer would likely have understood that his or her reported telescope observations of white dwarf stars would be relied upon for judging the safety of humanity's most ambitious particle experiment. Had astronomers been put on notice that their research results would be used as a foundation for real-world engineering safety analyses, it is quite possible they would have undertaken different or greater efforts to ensure reliability of the data.

***Errors in models, assumptions, and conceptual thinking***—Also important is the potential for mistakes in the reasoning process of science—building models, making appropriate assumptions, and guarding against conceptual error. When trying to describe the physical world and predict its behavior, it is always necessary to make assumptions, use models, and make the analysis manageable by omitting the consideration of various factors. For example, to determine how long it would take a car to brake to a stop on dry pavement, a researcher would omit to take into account the effects of special relativity. Such an analytical tack is untroubling. But in more complicated scientific questions, there are potential pitfalls in deciding which models to use, how complex the models must be, and what can be safely assumed.

---

581.   *Id.*

582.   *Id.* at 8.

583.   *See* LINCOLN, *supra* note 48, at 17. In discussing the relative porosity of solid matter to a fast-traveling neutral particle, Lincoln wrote, "Although the penetrating power of neutrinos depends somewhat on their energy, neutrinos of the energies typically seen in radioactive decay could pass through five light-years of solid lead with just a 50% probability of being detected. Five light-years equals more than 48 million kilometers, or 30 million miles." *Id.* In fact, five light years is approximately 47.3 trillion kilometers or 29.4 trillion miles. Thus Lincoln was off by six orders of magnitude and, even after the order of magnitude is adjusted, made a rounding error by saying "more than" instead of "slightly less than" or the equivalent.

584.   See CERN Scientific Policy Committee, *supra* note 232, at 2–3.

585.   Giddings & Mangano, *supra* note 211, at 23.

One example of incomplete modeling, pointed out by Ord, Hillerbrand, and Sandberg, is the Castle Bravo nuclear test of 1954.[586] The thermonuclear "H-bomb" device used a new kind of fusion fuel—lithium-6.[587] The lithium in the bomb was enriched to 40% lithium-6 isotope, with the remainder being lithium-7.[588] In predicting bomb yield, Los Alamos scientists used a model that did not include the decay of lithium-7.[589] As it turned out, lithium-7 reacted in unanticipated ways to neutron collisions, something for which the scientists were unprepared.[590] Within the nuclear reaction, when a lithium-7 nucleus was hit with a neutron, the lithium-7 nucleus expelled two neutrons in return, furthering the chain reaction.[591] The net loss of a neutron also converted the lithium-7 atom into lithium-6, which then provided additional fusion fuel.[592] As a result, instead of the predicted five-megaton yield, Castle Bravo exploded with 15 megatons, making it the largest yielding bomb in U.S. history.[593] Scientists were trapped in their experiment bunkers and ships placed outside the computed danger zone were forced to deal with Bravo's wrath.[594] A Japanese fishing vessel trawling outside the area cordoned off by the Navy was hit with fallout, killing one crew member.[595]

A subsequent thermonuclear device test, *Castle Romeo*, also created a runaway fireball.[596] That test yielded 11 megatons—triple the predicted yield—because of the same erroneous assumptions.[597]

Could the theoretical work done for the LHC safety argument be as flawed as the work of the Los Alamos scientists? There are reasons to think it might be. The Giddings and Mangano paper[598] makes many assumptions. While presumably reasonable on their face, many of these assumptions are untested. For instance, in determining that under certain scenarios black holes would grow too slowly to be a threat, Giddings and Mangano constructed a model in which black holes are assumed only to consume protons and neutrons, not electrons.[599] I am not specifically questioning the appropriateness of this or any other of their many assumptions. But it must be borne in mind that sometimes assumptions that seem reasonable at first turn out not to be so. Thus, a court faced with a case such as the LHC injunction should consider, in bulk, the quantity and kind of assumptions made in the analysis and the complexity of the argument. These factors are thus relevant considerations in making a rough measure of the strength of a scientific paper's overall conclusions.

---

586.   *See* Ord et al., *supra* note 543, at 9.
587.   *See* RICHARD RHODES, DARK SUN: THE MAKING OF THE HYDROGEN BOMB 541 (1995).
588.   *Id.*
589.   *Id.*
590.   *Id.*
591.   *Id.*
592.   *Id.*
593.   *Id.*
594.   *See id.*
595.   *Id.* at 542.
596.   *Id.*
597.   *Id.*
598.   *See generally* Giddings & Mangano, *supra* note 211.
599.   *Id.* at 8.

Looking specifically at the facts of the LHC case, it is worth considering that Rainer Plaga's paper[600] might well be the tip of the iceberg in terms of critiquing the models, methods, and conclusions of Giddings and Mangano. It seems to be the case that few people have the combination of training and incentive needed to undertake the detailed work of putting the Giddings and Mangano piece through its paces. If so, this cuts strongly against using the Giddings and Mangano paper as the sole foundation for judging that the LHC poses no risk whatsoever.

Moreover, work done a few years ago provides precedent suggesting that it is not disingenuous to second guess the work of contemporary physicists in this manner.

Cambridge physicist Adrian Kent pointed out in 2003 that the two papers written to address safety concerns about a potential strangelet disaster at the RHIC prior to its start contained rather striking conceptual math errors.[601] Three theoretical physicists from CERN, Arnon Dar, A. De Rújula, and Ulrich Heinz, wrote a paper called "Will Relativistic Heavy-ion Colliders Destroy Our Planet?" and made, in their words, the "extremely conservative conclusion [that] it is safe to run RHIC for 500 million years."[602] Kent pointed out that Dar, De Rújula, and Heinz misapprehended the nature of their own calculations.[603] Assuming them to be correct, the calculations meant only that there was a low probability that Earth would be destroyed very early on in such a run.[604] In fact, the CERN team's calculations were consistent with a high probability of planetary destruction over such a hypothetically long run.[605]

There was also error in the other major paper speaking to RHIC safety issues,[606] *Review of Speculative 'Disaster Scenarios' at RHIC.*[607] To recapitulate,[608] this paper was the result of a request of the Brookhaven lab's director to a committee of four physicists: Busza, Jaffe, and Wilczek of MIT and Sandweiss of Yale.[609] The authors commented that the Dar paper provided a ceiling on risk that was 100 million times the required safety margin.[610] However, as Kent pointed out, this would imply that the RHIC would be "safe" even if it had a high probability of destroying the Earth after just five years of operation.[611]

The Busza team did its own independent analysis, concluding that under one set of assumptions, the probability of a catastrophe was less than $2 \times 10^{-4}$, which they astonishingly described as leaving "a comfortable margin of error."[612] The numerical quantity is more meaningful when taken out of scientific notation: $2 \times 10^{-4}$ is just an alternative way of writing *one in 5,000*. That is a level of risk that few average people would regard as "comfortable" when the downside is crushing the Earth down to a stadium-sized ball of strange matter.

When Kent pointed out the analytical problems to the Busza team, the team revised their paper.[613] In their revised paper, they refigured the probabilities to add an extra factor of ten, thus claiming that there is no worse than a 1 in 50,000 risk of destroying the Earth—under their most conservative assumptions.[614] The Busza team also retreated on their policy pronouncements, writing, "We do not attempt to decide what is an acceptable upper limit on [the probability of a disaster], nor do we attempt a 'risk analysis,' weighing the probability of an adverse event against the severity of its consequences."[615] The Busza team cordially thanked Kent in the acknowledgments section of the revised document.[616]

Certainly everyone can make mistakes—even in printed articles. But the Busza article was no ordinary piece of scholarship. It was advanced as the basis upon which the public should feel at ease with the pending start up of the RHIC. With the article's real-world implications, and the fact that it had multiple authors and was reviewed by RHIC administrators for purposes of using it as a public-relations tool, the fallacious reasoning is astounding.

Considering the conceptual errors found in the Dar and Busza papers, and the fact that those papers were relied upon for safety judgments about the RHIC, one must take seriously the possibility that the Giddings and Mangano paper contains material errors—not yet uncovered—that would undermine or alter its bottom-line conclusion. It is precisely in this manner that the litigation process could be so valuable—fleshing out the issues through discovery and testimony, thus providing a judge with the chance to determine if serious questions exist.

---

610. *See id.* at 1126.

611. Kent, *supra* note 601, at 161.

612. *Id.*

613. *See* Jaffe, *supra* note 607, at 1125 n.1, 1126 n.2.

614. *See id.* at 1125, 1126.

615. *Id.* at 1126.

616. *Id.* at 1126 n.2, 1139.

### C.  *Analyzing the Potential for Credulity and Neglect*

The third mode of analysis that can be used in the LHC case to test for the existence of "serious questions" and *Harris-Stanley*-type negligence involves looking at the psychological and sociological factors at play. Does the social environment at CERN and in the particle-physics community encourage a full vetting of safety issues? Or, might concerns about safety be dismissed as fringe thinking, the expression of which becomes stifled? How might cognitive psychological effects shape the thinking and decision-making about potential risks? Is it plausible that scientists holding strong convictions in the LHC's safety might be engaging in subtle forms of self-deception? How might the development of people's opinions on matters of safety be affected by the fact that they have helped develop the LHC to this point? How might it affect their thinking that the LHC is so far along that the decision to operate it seems irreversible?

This section focuses on mistake without malice and how the courts can use discovery and evidence-gathering to determine whether it is plausible that scientists, as individuals and in groups, might be susceptible to errors in judgment even when their intentions are pure.

An essential step to understanding the vulnerability of the scientific process to these sources of error is realizing that science is a human enterprise, subject to human fallibility.

"There is a popular misconception that science is an impersonal, dispassionate, and thoroughly objective enterprise," wrote acclaimed physicist Paul Davies.[617] "This is, of course, manifest nonsense. Science is a people-driven activity like all human endeavor, and just as subject to fashion and whim."[618]

Beginning with the understanding that science is a human activity and scientists are humans, it is clear that a number of paradigms from psychology and sociology may shed light on how it might be plausible that scientists may unwittingly build an experimental apparatus that threatens humanity.[619]

***Cognitive dissonance***—The concept of cognitive dissonance may explain why particle physicists could deceive themselves into thinking that an experiment is not dangerous. Social psychologist Leon Festinger of Stanford University used the label "cognitive dissonance" to describe a distressing mental state in which people "find themselves doing things that do not fit with what they know, or having opinions that do not fit with other opinions they hold."[620] For example, the idea that someone had spent a large part of their career building a real-life doomsday machine would be dissonant with their self-view of being a caring person with a good heart.

---

617.  Paul Davies, *Introduction* to RICHARD P. FEYNMAN, SIX EASY PIECES: ESSENTIALS OF PHYSICS EXPLAINED BY ITS MOST BRILLIANT TEACHER ix (1995).

618.  *Id.*

619.  *See id.*

620.  *See* LEON FESTINGER, A THEORY OF COGNITIVE DISSONANCE 4 (1957); *see also* EM GRIFFIN, A FIRST LOOK AT COMMUNICATION THEORY 206, *available at* http://www.afirstlook.com/docs/cogdiss.pdf (2006) (quoting Festinger).

Moreover, the magnitude of the dissonance people feel increases with the importance of the issue and the amount of discrepancy between actions and privately held beliefs.[621] Thus, we would expect maximal dissonance in the case where the threatening cognition is a realization that one has been working toward the destruction of humanity.[622]

In addition, the more difficult it is to reverse a decision once made, the greater the resulting dissonance and the greater the need for reassurance.[623] At this point, halting the LHC would seem to be a practical impossibility since it represents sunk costs in the billions of dollars and the unrecoverable investment of large portions of many scientists' careers. Thus, the pressure from cognitive dissonance to refuse to believe that the LHC could put the world at risk would be predicted to be enormous.

In addition, cognitive dissonance can affect beliefs by indirect means of information filtering. Specifically, according to Festinger, people will seek information that is consistent with their beliefs and avoid information that is not.[624] Likewise, people tend to seek the company of like-minded people, thus buffering themselves from dissonance-causing information and ideas.[625]

Following on the work of Festinger, University of California social psychologist Elliot Aronson explained that humans are not rational animals, but *rationalizing* animals who seek to see themselves as reasonable.[626] With the specter of a planet-consuming black hole being so horrendous, the psyches of some scientists might simply reject the possibility—regardless of what the evidence is telling them. In this sense, it is interesting to recall that Einstein refused to believe in the existence of black holes.[627]

Psychologists have found that when a person is forced to do something contrary to his or her private opinion, under certain circumstances, that privately held opinion can actually change to more closely correspond with the person's overt behavior.[628]

Discovery and trial testimony could unmask the effects of cognitive dissonance. Questioning could reveal what people thought when, and might help to uncover beliefs based on self-deception.

***Reputational cascades***—Reputational cascades are phenomena where perceptions of risk build upon themselves as people tend to minimize risk in order earn social approval and avoid social disapproval.[629] Cass Sunstein points out two examples that illustrate how experts may be made to censor themselves: One medical researcher who was skeptical of a number of diagnoses of Lyme disease told the *New York Times,* "Doctors can't say what they think anymore . . . . If you quote me saying these things, I'm as good as dead."[630] And a sociologist who challenged widely held beliefs about mad-cow disease explained that by making such doubts known, "You get made to feel like a pedophile."[631]

---

621.  GRIFFIN, *supra* note 620, at 206.

622.  *See generally id.*

623.  *Id.* at 208.

624.  *Id.* at 207.

625.  *Id.*

626.  *Id.* at 211 (emphasis added).

627.  *See* SUSSKIND, *supra* note 153, at 30.

628.  *See* Leon Festinger & James M. Carlsmith, *Cognitive Consequences of Forced Compliance*, 58 J. OF ABNORMAL AND SOC. PSYCHOL. 203, 209 (1959).

One can imagine that in the small community of particle physicists, speaking out about LHC dangers could largely ruin someone's standing in the view of the community. It may be telling that Francesco Calogero, in his article critical of RHIC/strangelet risk assessment, thanked several colleagues without naming them, not wanting to compel them to be associated with views that the RHIC might be dangerous.[632]

***Confirmation bias***—Confirmation bias is the tendency of scientists to bolster hypotheses whose truth is in question.[633] Even when a scientist is engaged in an ostensibly unbiased effort to deduce truth, confirmation bias may unwittingly lead the scientist to selectively acquire and use evidence in a manner that supports the hypothesis.[634]

A great deal of empirical research in experimental psychology has shown that confirmation bias appears in many forms with strong effects.[635] Confirmation bias works from both sides of the process of reasoning and investigation. People tend to seek evidence that supports a hypothesis, and people tend to avoid evidence that would undermine a hypothesis.[636] Posner has noted that in areas of scientific uncertainty, career concerns can influence scientific judgments.[637]

Confirmation bias may help to explain what went wrong in the Space Shuttle *Columbia* disaster. The report of the Columbia Accident Investigation Board ("CAIB") found that decision makers focused on information that tended to support their expected or desired result—that the foam strike that ultimately doomed *Columbia* did not represent a safety of flight issue.[638]

---

Given what is on the line with the LHC, the desired result for particle physicists is clear.

In terms of confirmation bias, another possible parallel between the *Columbia* flight and the LHC has to do with keeping to schedules. The CAIB report placed considerable emphasis on the shuttle program's focus on keeping to a tight and unforgiving schedule and reported that mission managers tended to draw conclusions that minimized the risk of delaying future launches.[639]

One might imagine that similar pressures were on Giddings and Mangano, the LSAG, and the SPC, especially considering that they were doing their work in the last months before the LHC's start up. A less than completely confident endorsement of the LHC on safety issues might have delayed or even doomed the project.

The problems of confirmation bias are implicitly recognized in the NEPA legislative scheme, which requires environmental impact statements before projects are undertaken.[640] The point was underscored by Justice Stephen Breyer in an opinion concerning NEPA, written when he was a judge on the First Circuit: "The way that harm arises may well have to do with the psychology of decision makers, and perhaps a more deeply rooted human psychological instinct not to tear down projects once they are built."[641]

Although some might think that a "hard" science like physics would be immune from such bias, this is not so. Monwhea Jeng, a physicist at Southern Illinois University, has compiled a set of historical examples showing how the beliefs of physicists have influenced their results.[642] While it may be easy to believe expectation bias could occur in modeling and choosing assumptions, expectation bias can even infect purely mathematical operations. Jeng relays an example from particle physics where researchers undertook complex calculations in the search for free quarks.[643] The calculations matched up

---

extremely well with the expected factional charges of quarks.[644] But when the calculations were reworked on a "blind" basis—with the data subjected to a random offset that was only removed at the end after all calculations were complete, the results did not line up with models of quark charge at all.[645]

***Groupthink***— In trying to take seriously the question of whether the LHC really does present a planet-threatening danger, we are presented with a paradox. How could so many exceedingly smart people—particle physicists, no less, who make a cliché of genius—be capable of getting something so terribly wrong?

Psychologist Irving L. Janis worked up his theory of groupthink when trying to resolve a paradox of his own.[646] Reading a history on the 1961 Bay of Pigs fiasco, when U.S. operatives staged a spectacularly failed invasion of Cuba, Janis found himself asking, "How could bright, shrewd men like John F. Kennedy and his advisers be taken in by the CIA's stupid, patchwork plan?"[647] His answer was "groupthink."[648] The groupthink concept acknowledges that "smart people working collectively can be dumber than the sum of their brains."[649] According to Janis, groupthink is "a mode of thinking that people engage in when they are deeply involved in a cohesive in-group, when the members' strivings for unanimity override their motivation to realistically appraise alternative courses of action."[650] It involves "a deterioration of mental efficiency, reality testing, and moral judgment that results from in-group pressures."[651]

The effect can be insidious. The process allows individuals to maintain a worry-free outlook that is not justified by the facts. In such a dynamic, the existence of group consensus causes individuals to forego or dismiss their own independent thinking. A circularity develops: Group consensus justifies individual confidence, and individual confidence justifies group consensus. The result is flawed decision-making.

The classical antecedents for groupthink are group cohesiveness, leadership's existing preference for a certain decision, and insulation of the group from outside opinions.[652] Considering these factors, CERN would appear highly vulnerable to groupthink.

---

644.  *Id.*

645.  *See id.*

646.  IRVING L. JANIS, VICTIMS OF GROUPTHINK: A PSYCHOLOGICAL STUDY OF FOREIGN-POLICY DECISIONS AND FIASCOES iii (1972).

647.  *Id.*

648.  *See id.* at iii, 9.

649.  John Schwartz & Matthew L. Wald, *NASA's Curse?: 'Groupthink' Is 30 Years Old, and Still Going Strong*, N.Y. TIMES, Mar. 9, 2003 § 4, at 5.

650.  JANIS, *supra* note 646, at 9.

651.  *Id*.

652.  *See* Gregory Moorhead, Richard Ference, & Chris P. Neck, Group Decision Fiascoes Continue: Space Shuttle Challenger and a Revised Groupthink Framework, 44 HUM. REL. 539, 541 (1991); *see also* JANIS, *supra* note 646, at 197.

*Group cohesiveness*—Anthropologist Sharon Traweek, in an ethnography on particle physicists, wrote that the physics field forms "an extremely restricted community."[653] The most important communications are made by word of mouth, through means such as informal talks and seminars, with papers being less important.[654] Those particle physicists who do not know each other well, want to.[655]

"Physicists seem to be in constant circulation, moving around the world from lab to lab, department to department, always talking, forming alliances and collaborations. Most important, they are bound together by a way of thinking, about the world and about knowledge and about themselves."[656]

*Leadership's existing preference for a certain decision*—This aspect of the particle-physics community, especially within CERN, seems very clear: The preferred outcome is a determination that the LHC is safe. In this respect, it should be remembered that the LHC is not a small part of CERN's research program. In an agreement with the U.S. Department of Energy, CERN confirmed in a recital the "overriding priority and the vital importance of the LHC for the future of the laboratory."[657] Without the LHC, CERN would be reduced to nearly nothing.[658]

*Insulation of the group from outside opinions*—According to Traweek, particle physicists "construct their world and represent it to themselves as free of their own agency."[659] Theirs is "an extreme culture of objectivity: a culture of no culture, which longs passionately for a world without loose ends, without temperament, gender, nationalism, or other sources of disorder—for a world outside human space and time."[660]

Discussing the Stanford Linear Accelerator Center ("SLAC"), Traweek noted that "disagreement with the lab's policies is seen as a result of a lack of information, which SLAC will supply."[661] And in describing the process of training new physicists, Traweek reported, "At the major labs, they learn that outsiders are devalued and exactly how this is done and what justifications are given."[662]

Traweek also described the "tremendous force of the division in the physicists' cosmology between outsiders, no matter how well-informed, and

---

insiders."[663] In the academic realm, particle physicists think particle physics requires the most intelligence and reasoning capacity, and humanities the least.[664]

Along these lines, it is interesting to consider the marginalization that Rainer Plaga—an astrophysicist by training—received in response to his article politely questioning the Giddings and Mangano paper.[665]

If a groupthink dynamic develops, Janis describes several symptoms that can lead to flawed decision-making, including: (1) "an illusion of invulnerability, shared by most or all the members, which creates excessive optimism and encourages taking extreme risks," (2) "collective efforts to rationalize in order to discount warnings," (3) "an unquestioned belief in the group's inherent morality, inclining the members to ignore the ethical or moral consequences of their decisions," (4) "stereotyped views of" the enemy as "evil," "weak," or "stupid," (5) "direct pressure on any member who expresses strong arguments against any of the group's stereotypes, illusions, or commitments, making clear that this type of dissent is contrary to what is expected of all loyal members," (6) "self-censorship of deviations from the apparent group consensus, reflecting each member's inclination to minimize to himself the importance of his doubts and counterarguments," and (7) "a shared illusion of unanimity concerning judgments conforming to the majority view (partly resulting from self-censorship of deviations, augmented by the false assumption that silence means consent)."[666]

Court-compelled discovery would be an ideal vehicle for exploring the extent to which these symptoms are evident within the CERN community. For example, depositions and oral testimony could uncover examples of self-censorship and shared illusions of unanimity.

But even without discovery, for some of these symptoms, there is evidence to be found from publicly available sources. With regard to a disparaging characterization of enemies, recall that Ellis painted detractors as "nuts" and engaging in criticism for money[667] and that Brian Cox used a vulgar slur to refer to people who gave credence to the black-hole scenario.[668]

As to having an unquestioned belief in the group's inherent morality, Traweek notes:

> They have a passionate dedication to this vision of unchanging order: they are convinced that the deepest truths must be static, independent of human frailty and hubris. Simultaneously, they believe that this grand structure of physical

---

663.  *Id.* at 14.

664.  *See id.* at 79.

665.  *See* discussion *supra* Part IV.H.

666.  Janis, *supra* note 646, at 197–98.

667.  Ellis, *supra* note 83 (beginning at 38, 49 min.).

668.  *See* O'Neill, *Anyone Who Thinks the LHC Will Destroy the World is a Twat*, *supra* note 356.

truth can be progressively uncovered, and that this is the highest and most urgent human pursuit.[669]

In terms of exerting direct pressure on members of the group who do not fall in line, a perhaps telling anecdote comes from Ellis's talk at CERN about LHC/disaster issues.[670] Holding a laser pointer for his presentation, Ellis asked people to put their hands up if they believed that, if microscopic black holes occur, they would be stable.[671] Then, Ellis chided the audience, "Don't forget, I got this laser here."[672]

While he said this with a laugh, the message seemed to be clear—that dissent would not be looked upon kindly. "I remind you," he continued, "that you expect them to decay because of Hawking radiation."[673]

### D. Analyzing the Potential for Bias and Influence

The fourth category of analysis for a court in a case such as this involves looking at the potential for bias among those who have made pronouncements about safety, and the potential for influence to be brought to bear on that assessment process. The propriety of such an inquiry is clear, and it is closely tied to the historical role of judges and juries in evaluating evidence by using indicia of trustworthiness and by patiently listening to opposing parties' attempts to impeach the credibility of witnesses.

By way of previewing this analysis for the LHC case, I will say much less than I could. The public record of the controversy reveals for plaintiffs an embarrassment of riches in arguments to show bias among the assessors.

It is remarkable to think for a moment how CERN's situation might be viewed if, instead of operating a particle accelerator, CERN was a developer of pharmaceuticals. If a pharmaceutical firm attempted to take a drug to market based on the safety assessment of a panel of five of its employees, who in turn relied on the scientific work of one employee and one other scientist with a pending visiting position with the firm—it would be a scandal of epic proportions.

In the small field of particle physics where everyone seems to know everyone,[674] to demonstrate true independence, scientists rendering views on safety need more than the simple absence of formal institutional affiliations with the labs. Yet, the scientists rendering opinions on LHC safety lack even that. Indeed, every major paper relied upon for demonstrating the safety of the RHIC and the LHC has had a compromising institutional affiliation. For one paper on the RHIC, three of the four authors were experimental physicists planning to conduct experiments at RHIC, who then went on to do so.[675] The other paper on RHIC safety was written by three theorists at CERN, whose ALICE experiment on the LHC would be subject to the same objections with regard to strangelet dangers.[676] The 2003 report on LHC safety, concerning the heavy-ion ALICE experiment, while the least dominated by institutional insiders, still included a CERN scientist on the panel,[677] even though CERN labeled the authors as a "group of independent scientists."[678] Moreover, the two major papers relevant to the current safety concerns—those regarding proton-proton collisions and black holes—do not include a single non-CERN-affiliated author.

---

669.   TRAWEEK, *supra* note 247, at 17.
670.   Ellis, *supra* note 83.
671.   *Id.* (beginning at 35 min.).
672.   *Id.* (beginning at 35 min.).
673.   *Id.* (beginning at 35 min.).
674.   TRAWEEK, *supra* note 247, at 3.

We must ask if there is something special about particle physics that makes conflicts of interest untroubling. Indeed, CERN physicists seem to proceed as if conflicts of interest are immaterial. That much is implied by the open and unhidden nature of the conflicted interests found in LHC safety assessments. Yet the authors offer no arguments for why they ought to be exempt from the usual norms of establishing trustworthiness by avoiding self-dealing.

In short, there is no reason not to accept the obvious: Conflicts of interest in judgments about particle physics are just as troubling as they are in any field.

At least some physicists seem to agree with this assessment. Adrian Kent of Cambridge wrote:

> Future policy on catastrophe risks would be more rational, and more deserving of public trust, if acceptable risk bounds were generally agreed upon ahead of time and if serious research on whether those bounds could indeed be guaranteed was carried out well in advance of any hypothetically risky experiment, with the relevant debates involving experts with no stake in the experiments under consideration.[679]

Likewise, Francesco Calogero wrote that "it is of course appropriate that, to the maximum extent possible, those who are assigned the task of making such evaluations should not be affected by any conflict of interest."[680] Calogero specifically recommended a "blue team" performing an analysis that is as objective as possible, and a "red team" working as devil's advocates trying to prove that the experiment is dangerous.[681]

As a counter to these sorts of suggestions, LHC defenders sometimes pose this rhetorical question: How could particle physicists support a course of action that would destroy the Earth, given that they love their children and

---

value their own lives? The problem with the argument implied by this question is that it sets up a non-existent dilemma. Clearly, particle physicists are not intentionally attempting to destroy the Earth, nor are physicists engaging in a course of action where the probability of destroying the Earth is more likely than not. Instead, the issue of LHC risk is one of low probabilities—risks such as one in 5,000, one in 100,000, one in a million, or even lower. Thus, the relevant question to ask is, would particle physicists tolerate a higher risk of catastrophe from their experiments than the rest of humanity might?

The answer is quite obvious. Of course they might. And that answer is not mere speculation. Physicists have actually published a hard number: Busza, Jaffe, Sandweiss, and Wilczek regarded a one-in-5,000 risk of destroying Earth as "comfortable."[682]

It seems manifest that most people—that is, those having no involvement in physics experimentation—would not regard a one-in-5,000 risk of the destruction of Earth as a comfortable risk to assume for a science experiment—especially one conducted for purely academic reasons.

Keeping this precedent in mind, it is appropriate to scrutinize the bottom-line conclusions of Giddings and Mangano. Those conclusions were that there is "no basis for concerns" about LHC-produced black holes and that "there is no risk of any significance whatsoever from such black holes."[683]

What level of risk is significant to particle physicist with an interest in the experiments, and what level of risk is significant to an uninterested party, will undoubtedly be different. Thus, the ultimate conclusion of Giddings and Mangano's work that there is "no risk of any significance" should be treated skeptically as a value judgment—that is, a subjective ethical pronouncement—rather than as a hard scientific conclusion.

On the basis of the above, I think one must conclude that it is plausible that particle physicists would be quite biased when it comes to assessments for low-probability risks. But let's take our analysis further and explore whether it is plausible that particle physicists might be biased when it comes to higher probability risks—something like one in five or one in twenty. There appears to be no indication that the actual risk posed by the LHC might approach such appalling odds. But for the sake of completeness, let's consider, hypothetically, whether it is plausible that particle physicists might be willing to gamble the fate of the world under those circumstances.

To begin with, it seems highly likely that particle physicists might fear serious reprisals and negative repercussions for their careers if they were to speak out about perceived dangers of the LHC. An academic in such a situation—even one not affiliated with CERN—might plausibly face denial of tenure, unaccepted manuscripts, and ostracism by peers.

In mulling over whether to speak out, particle physicists with private doubts might well resign themselves to a fatalistic assessment. They might plausibly figure that they, as individuals, are powerless to overcome the momentum of a multinational multi-billion-dollar project. If that is their appraisal, then such individuals have nothing to gain, but much to lose, by making a public objection. Consider the possible outcomes: If a scientist speaks out and nothing bad happens, the scientist is a laughingstock. If a scientist speaks out and disaster does come to pass, professional vindication will be fleeting and bittersweet. If a scientist keeps mum or even extols the safety of the project, in a disaster scenario, embarrassment will be short-lived.

---

682.  See Kent, *supra* note 601, at 161.
683.  Giddings & Mangano, *supra* note 211, at 27.

On the other hand, suppose particle physicists with private doubts reach the opposite conclusion about the likely impact of their public dissent. Suppose a private doubter predicts that his or her voice could be the tipping point that leads to widespread public concern and a permanent shutdown of the LHC. In such a case, whether the objecting scientist is right or wrong, he or she can anticipate being blamed for ruining the most exciting opportunity for advancing fundamental physics in a generation. And there's no hope of vindication in such an event: Naysayers cannot be proved right if the experiments are never run.

The math-oriented are often fond of using matrices to elucidate decision-making. A physicist creating such a matrix, using the logic detailed above, would be faced with a series of boxes in which all outcomes are quite bad, except one: to be a supporter of the LHC in the event that it turns out to be a benign scientific triumph. Even if one does so with fingers crossed for good luck.

## VIII.  CONCLUSION

Am I fearful that the world will be destroyed by a lab-created black hole? No. Not really. It does not keep me awake at night. All things considered, the odds seem quite slim.

But a seemingly slim possibility, even if it leaves me personally unanxious, does not indicate that the risk is insignificant. Even a tiny chance of a black-hole catastrophe could be very significant as matter of equity before a court. The alleged downside, after all, is the disappearance of our planet down a cosmic drain.

From my perspective as a lawyer, sizing up the merits of the case, I find the assurances provided by the particle-physics community to be quite lacking. In particular, I am struck by the fact that the safety assurances are based on scientific work that brazenly lacks independence.

I have no reason to doubt that particle physicists are, generally speaking, good people. But I am not at all convinced that the field is so pristine, so cleansed of human foible, that it has no need for the ordinary indicia of verisimilitude.

More specifically, the history of the black-hole debate leaves me uneasy. There is a repeating pattern of airtight assurances—presented with utter conviction—that are quietly abandoned later when the scientific bedrock upon which they are based suddenly erodes. The latest argument, unveiled in 2008 on the eve of the planned start up of the LHC, was offered with the usual sense of resolute confidence. But one naturally wonders whether history will repeat itself and time will reveal new scientific understanding that trumps the seemingly incontrovertible assumptions underlying this latest assurance of safety.

All that being said, let me reiterate—the odds of doomsday seem very slight, even accepting most of the arguments of the LHC's critics. But slim odds must be weighed against the grim downside.

My motivation in writing is certainly not to engender fear. I have no apprehension to share. Nor is it my intent or my desire to shut down the LHC. Mine is not a policy argument. Frankly, my research for this article has intensified my armchair interest in seeing what results from the LHC's novel experiments. My argument is one of law. My conviction is that, when a black-hole case arrives on a docket, no court should abdicate its role as a bursar of equity, even where, as here, the socio-political pressure to abstain will be immense, the factual terrain will be intensely intellectually challenging, and the jurisprudential conundrums are legion. At the end of the day, whether the LHC represents an intolerable danger is, in my view, an open question. I have not endeavored to provide a definitive answer. But I think the courts should.

It is part of our 21st Century reality that we must take seriously a number of surreal planetary disaster scenarios. In that sense, the synthetic-black-hole disaster is not unique. For some time now, we have been confronted with the possibility of nuclear war and global climate change. In the future, we may have to remove still more scenarios from the science fiction category and place them on a list of real worries. Someday, we may need to seriously consider catastrophic threats from nanotechnology, genetic engineering, or artificial intelligence. Each one of these human-made global disaster scenarios involves incredibly complex questions of science, engineering, and mathematics. Courts must develop tools to deal meaningfully with such complexity. Otherwise, the wildly expanding sphere of human knowledge will overwhelm the institution of the courts and undo the rule of law—just when we need it most.